\documentclass[10pt,twocolumn,aps,amsmath,superscriptaddress,amssymb]{revtex4-1}
\usepackage[latin9]{inputenc}
\usepackage{mathtools}
\usepackage{amsmath}
\usepackage{amssymb}
\usepackage{cancel}
\usepackage{mathdots}
\usepackage{stmaryrd}
\usepackage{stackrel}
\usepackage{graphicx}
\PassOptionsToPackage{version=3}{mhchem}
\usepackage{mhchem}
\usepackage{esint}

\makeatletter

\usepackage{braket}
\usepackage{dsfont}

\renewcommand{\vec}{\mathbf}

\renewcommand{\Re}{\mathop{\mathrm{Re}}}
\renewcommand{\log}{{\,\mathrm{ln}}}                      

\makeatother

\begin{document}

\title{Controlling competing orders via non-equilibrium acoustic phonons:
emergence of anisotropic effective electronic temperature}

\author{Michael Schütt}

\affiliation{School of Physics and Astronomy, University of Minnesota, Minneapolis,
Minnesota 55455, USA}

\author{Peter P. Orth}

\affiliation{Department of Physics and Astronomy, Iowa State University, Ames,
Iowa 50011, USA}

\author{Alex Levchenko}

\affiliation{Department of Physics, University of Wisconsin-Madison, Madison,
Wisconsin 53706, USA}

\author{Rafael M. Fernandes}

\affiliation{School of Physics and Astronomy, University of Minnesota, Minneapolis,
Minnesota 55455, USA}
\begin{abstract}
Ultrafast perturbations offer a unique tool to manipulate correlated
systems due to their ability to promote transient behaviors with no
equilibrium counterpart. A widely employed strategy is the excitation
of coherent optical phonons, as they can cause significant changes
in the electronic structure and interactions on short time scales.
One of the issues, however, is the inevitable heating that accompanies
these resonant excitations. Here, we explore a promising alternative
route: the non-equilibrium excitation of acoustic phonons, which,
due to their low excitation energies, generally lead to less heating.
We demonstrate that driving acoustic phonons leads to the remarkable
phenomenon of a momentum-dependent effective temperature, by which
electronic states at different regions of the Fermi surface are subject
to distinct local temperatures. Such an anisotropic effective electronic
temperature can have a profound effect on the delicate balance between
competing ordered states in unconventional superconductors, opening
a novel avenue to control correlated phases. 
\end{abstract}

\date{\today}

\maketitle

\section{Introduction}

One of the hallmarks of correlated electronic systems is the existence
of multiple electronic orders with comparable energy scales that entangle
different degrees of freedom. In the case of unconventional superconductors,
such as iron pnictides, cuprates, and heavy fermions, superconductivity
is usually found to compete with a density-wave type of order, characterized
by a non-zero wave-vector $\mathbf{Q}$~\cite{MoonPRB2010}. For
instance, in the pnictides, $s^{+-}$ superconductivity competes with
$\mathbf{Q}=\left(\pi,0\right)$ spin density-wave order~\cite{FernandesPRB2010};
in the cuprates, the $d$-wave superconducting transition temperature
$T_{c}$ is suppressed by the onset of incommensurate charge order
with ordering vector $\mathbf{Q}\approx\left(\frac{\pi}{3a},0\right)$~\cite{ChangNP2012};
the superconducting state of the heavy fermions, on the other hand,
competes with a Néel-type magnetic order, characterized by $\mathbf{Q}=\left(\pi,\pi\right)$~\cite{PhamPRL2006}.
These observations open up the interesting possibility of enhancing
superconductivity by suppressing the competing density-wave order.

The standard ways to experimentally tune the superconducting and density-wave
orders in these materials is by chemical substitution, pressure, or
magnetic field. Recent advances in ultrafast pump-and-probe techniques,
however, opened a new avenue to explore and control these competing
electronic states, as they generally undergo different relaxation
processes to return to equilibrium \cite{Orenstein-PhysToday-2012,Giannetti-AdvPhys-2016,MatteoN2007,Fausti14012011,YangPRL2014,MankowskyN2014,VishikArxiv2016,Mitrano2016,ChiaPRL2010}.
Importantly, by taking the system out of equilibrium, its parameters
change on ultrafast time scales, which can result in transient electronic
states with no equilibrium counterpart. Indeed, an earlier experimental
demonstration of non-equilibrium control of electronic order, based
on pioneering ideas by Eliashberg \cite{Eliashberg,EliashbergInLangenbergLargkin1986},
was the surprising enhancement of superconductivity by sub-gap microwave
irradiation~\cite{Pals-PhysRep-1989}. More recently, ultrafast perturbations
have been widely employed to manipulate unconventional superconductors
by selecting particular lattice excitations of the system \emph{via}
optical pulses resonant with optical phonon modes \cite{SubediPRB2014}.
These coherent lattice excitations then modify the electronic structure
and interactions of the system, which in turn can favor superconductivity
or other electronic ordered states \textendash{} a concept widely
explored theoretically \cite{FuPRB2014,MoorPRB2014,DzeroPRB2015,RainesPRB2015,WangPRL2016,KnapPRB2016,KennesArXiv2016,KemperArXiv2016,BabadiArXiv2017}.
Examples include the melting of stripe order in the cuprates \cite{Fausti14012011},
the coherent modulation of the chemical potential in the pnictides
\cite{YangPRL2014}, and the possible promotion of transient superconductivity
at high temperatures in organic superconductors \cite{Mitrano2016}.
There are however important issues intrinsic to this approach. First,
the fact that the transient light-induced states only exist on ultrashort
few picosecond time scales poses a key challenge to stabilize such
non-equilibrium states of matter for longer times. Second, the resonant
excitation of optical phonons inevitably leads to significant heating,
which generally suppresses the desired transient states \cite{Werner17}.

\begin{figure*}
\begin{centering}
\includegraphics[width=0.8\linewidth]{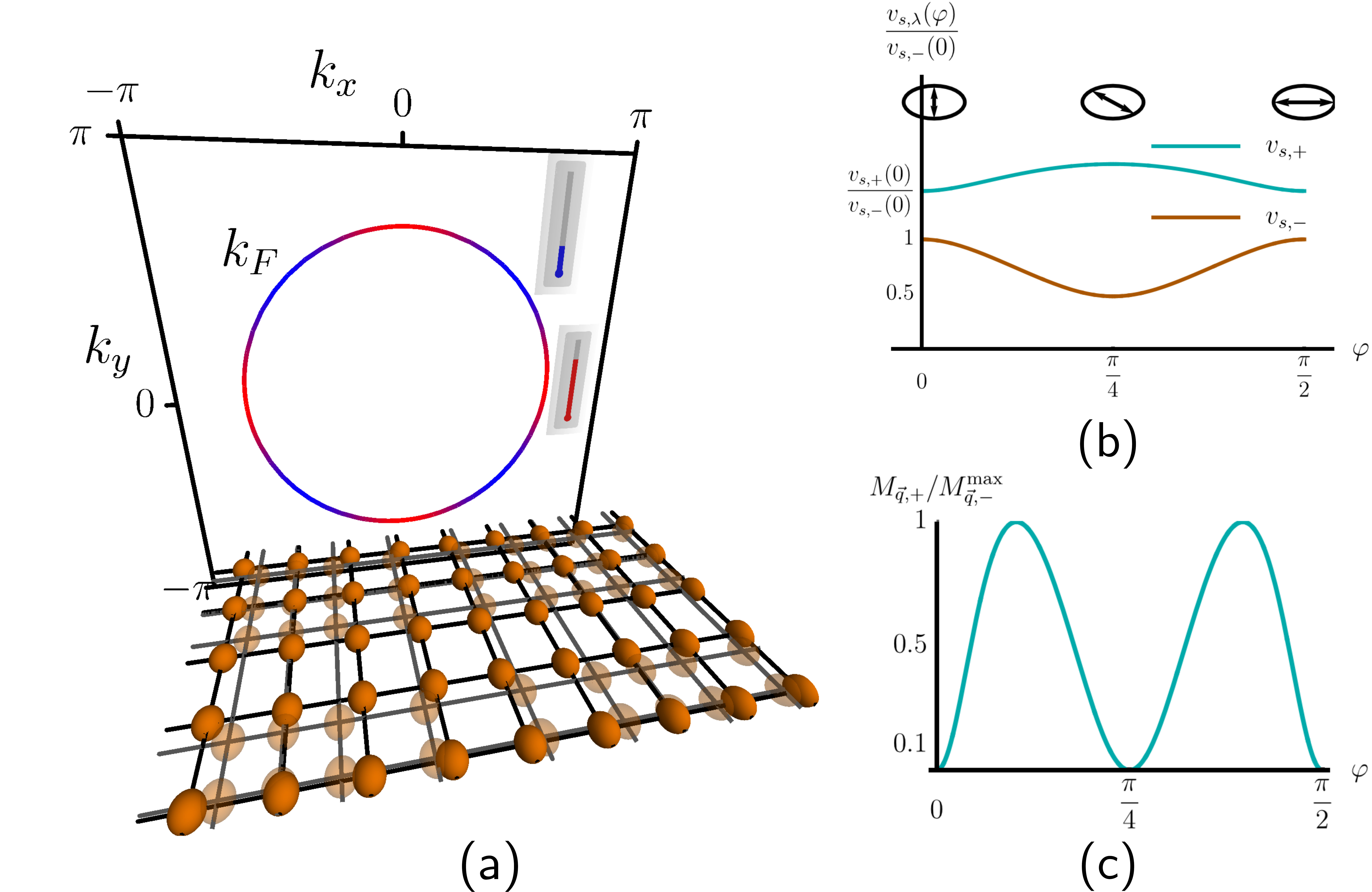} 
\par\end{centering}
\caption{\textbf{Momentum-dependent effective electronic temperature generated
by non-equilibrium acoustic phonons}. (a) The lower panel illustrates
the instantaneous distortions on a square lattice caused by the excitation
of an acoustic phonon mode. The solid (semi-transparent) lines and
symbols refer to the undistorted (distorted) lattice. Scattering by
non-equilibrium acoustic phonons promotes a momentum-dependent redistribution
of electronic quasi-particles, which is translated as a steady-state
effective temperature profile that varies along the Fermi surface,
depicted in the upper panel (blue and red represent local temperatures
that are colder and hotter than the average, respectively). (b) Anisotropic
sound velocity $v_{s,\lambda}$ of the two in-plane acoustic phonon
modes as a function of the propagation direction $\varphi$. The polarization
directions along three high-symmetry directions are also shown. (c)
Electron-phonon matrix element $M_{q,-}$ for the low-energy phonon
mode as a function of the propagation direction $\varphi$. In this
and in the next figures, we set the ratio $(C_{11}-C_{12})/(2C_{66})=1/4$, such that $v_{s,-}(\pi/4)/v_{s,-}(0) = 1/2$. }
\label{FigPhononLatticeDistortionIllustration} 
\end{figure*}

In this paper, we explore an alternative, complementary route to manipulate
competing electronic phases on long time scales: the non-equilibrium
excitation of acoustic phonons. Although acoustic phonons do not couple
directly to light, they can be excited by rapidly applying lattice
strain via an interface or via piezoelectric forces \cite{MatteoN2007,PezerilPRL2011,CavigliaPRL2012,ForstNM2015}.
The small energies required to excite acoustic phonons, as compared
to optical phonons, generate less heating. Importantly, while the
excited optical phonon modes are usually isotropic, the acoustic phonons
studied here are intrinsically anisotropic. As a result, the coupling
between the electronic subsystem and the non-equilibrium distribution
of acoustic phonons leads to a redistribution of electronic quasiparticles
close to the Fermi surface without generating too much heating. As
we show below, this redistribution of electronic spectral weight is
then translated into an \emph{anisotropic effective electronic temperature
profile} $\delta T_{\mathbf{p}}$ on the Fermi surface, resulting
in momentum-selective heating of the low-energy electronic states
(see schematic Fig.~\ref{FigPhononLatticeDistortionIllustration}a).
The term effective temperature is sharply defined below; for now,
we emphasize that it is not the thermodynamic temperature, but a parametrization
of the non-equilibrium distribution function that appears on the non-equilibrium
gap equations in a similar way as the thermodynamic temperature appears
on the equilibrium gap equations.

Our calculations reveal that the precise shape of the effective temperature
profile $\delta T_{\mathbf{p}}$ can be tuned by selecting the energy
of the excited acoustic phonons. Furthermore, because the anisotropy
of $\delta T_{\mathbf{p}}$ is a consequence of the anisotropy of
the phonon velocity, the effect discovered here is amplified near
a structural phase transition, where the phonon velocity is strongly
suppressed along particular directions. This feature makes iron pnictides,
cuprates, and heavy fermions promising systems to be manipulated by
non-equilibrium acoustic phonons, since some of their phase diagrams
display large nematic fluctuations, which cause strong lattice softening
\cite{FernandesNP2014,FradkinARCMP2010}.

The steady state characterized by a momentum-dependent effective temperature
is therefore fundamentally different from any equilibrium state, and
allows the control of competing electronic states, particularly in
the case of superconductivity competing with a density-wave type of
order. The main point is that while generally most low-energy electronic
states contribute to the superconducting instability, the competing
density-wave instabilities are mostly affected by particular points
of the Fermi surface \textendash{} the ``hot spots'' connected by
the non-zero ordering vector $\mathbf{Q}$. As a result, when the
effective local temperature of these hot spots is larger than the
average effective temperature across the entire Fermi surface, the
competing density-wave instability is suppressed, and superconductivity
is enhanced. We demonstrate this very general effect below by explicitly
calculating, using the Keldysh formalism, the steady-state phase diagram
of a low-energy model widely employed to study competing superconducting
and spin-density wave order in the iron pnictides. Finally, we also
discuss the conditions that are necessary for the effect discussed
here to not be washed away by other relevant relaxation processes,
such as those arising from impurity scattering and electron-electron
scattering.

\section{Microscopic model \label{sec_Model}}

Our starting point is the electron-phonon Hamiltonian $H=H_{\mathrm{el}}+H_{\mathrm{ph}}+H_{\mathrm{el-ph}}$.
Here, $H_{\mathrm{el}}$ describes an interacting electronic system
$H_{\text{el}}=H_{0}+H_{\text{int}}$ with kinetic term $H_{0}=\sum_{\mathbf{p}}\xi_{\mathbf{p}}c_{\mathbf{p},\sigma}^{\dag}c_{\mathbf{p},\sigma}$,
where the operator $c_{\vec{p},\sigma}^{\dag}$ creates an electron
with momentum $\vec{p}$ and spin $\sigma$ and $\xi_{\vec{p}}$ is
the band dispersion. The interaction term $H_{\text{int}}$ is responsible
for the superconducting and density-wave instabilities of the system,
and will be discussed in more details later. The phonons are described
by the harmonic term $H_{\mathrm{ph}}=\sum_{\vec{q},\lambda}\omega_{\vec{q},\lambda}a_{\vec{q},\lambda}^{\dagger}a_{\vec{q},\lambda}^{\phantom{\dagger}}$,
where the operator $a_{\vec{q},\lambda}^{\dag}$ creates a phonon
with momentum $\vec{q}$ and energy $\omega_{\vec{q},\lambda}$ in
branch $\lambda$. In the long wavelength regime, $\omega_{\vec{q},\lambda}=v_{s,\lambda}(\varphi_{\vec{q}})|\vec{q}|$,
where we introduced the sound velocity $v_{s,\lambda}$, which depends
on the propagation direction $\varphi_{\vec{q}}\equiv\tan^{-1}(q_{y}/q_{x})$.
Finally, the electron-phonon term is: 
\begin{equation}
H_{\mathrm{el-ph}}=\sum_{\substack{\vec{p},\vec{p}',\vec{q}\\
\vec{G},\lambda,\sigma
}
}\bigl(M_{\vec{p},\vec{p}',\vec{q},\lambda}\,a_{\vec{q},\lambda}^{\dagger}c_{\vec{p},\sigma}^{\dagger}c_{\vec{p}',\sigma}^{\vphantom{\dagger}}\delta_{\vec{p}'-\vec{p}-\vec{q}+\vec{G}}+\mathrm{h.c.}\bigr)\label{H_el_ph}
\end{equation}
with momentum conserved up to a reciprocal lattice vector $\vec{G}$.
In the absence of umklapp processes, the matrix element $M_{\vec{p},\vec{p}',\vec{q},\lambda}$
is typically approximated, in the long wavelength regime, by \cite{Ziman2001}
\begin{equation}
|M_{\vec{p}-\vec{p}'=\vec{q},\lambda}|^{2}\propto\frac{\left(\vec{q}\cdot\vec{e}_{\vec{q},\lambda}\right)^{2}}{v_{s,\lambda}(\varphi_{\vec{q}})|\vec{q}|}\label{matrix_element}
\end{equation}
with transferred momentum $\vec{q}$ and phonon polarization $\vec{e}_{\vec{q},\lambda}$.
The polarization and dispersion of the acoustic phonon modes are determined
solely by the properties of the elastic tensor. Since many of our
systems of interest are tetragonal, the elastic free energy is given
by $F=\sum_{\mathbf{q},ij}u_{i}\mathcal{D}_{ij}(\vec{q})u_{j}$, where
$\mathbf{u}$ is the lattice displacement and $\mathcal{D}_{ij}(\vec{q})$
is the symmetric dynamic matrix of a tetragonal system:
\begin{widetext}
\begin{equation}
\mathcal{D}_{ij}(\vec{q})=\begin{pmatrix}C_{11}q_{x}^{2}+C_{66}q_{y}^{2}+C_{44}q_{z}^{2} & (C_{12}+C_{66})q_{x}q_{y} & (C_{13}+C_{14})q_{x}q_{z}\\
(C_{12}+C_{66})q_{x}q_{y} & C_{66}q_{x}^{2}+C_{11}q_{y}^{2}+C_{44}q_{z}^{2} & (C_{13}+C_{44})q_{y}q_{z}\\
(C_{13}+C_{44})q_{y}q_{z} & (C_{13}+C_{14})q_{x}q_{z} & C_{44}(q_{x}^{2}+q_{y}^{2})+C_{33}q_{z}^{2}
\end{pmatrix}\label{dynamic_matrix}
\end{equation}
\end{widetext}

Here, $C_{ij}$ are the elastic constants. Because our focus is in
layered systems, we set $q_{z}=0$ hereafter. Instead of writing the
expressions in terms of the three relevant elastic constants $C_{11}$,
$C_{12}$, and $C_{66}$, it is convinient to introduce the three
coefficients $\mu_{1}=C_{11}+C_{66}$, $\mu_{2}=C_{11}-C_{66}$ and
$\mu_{3}=C_{12}+C_{66}$. Diagonalization of the symmetric matrix
leads to two phonon branches $\lambda=\pm$; their velocities are
given by:

\begin{equation}
v_{s,\pm}(\varphi_{\vec{q}})=\frac{1}{\sqrt{2\rho}}\sqrt{\mu_{1}\pm\sqrt{\frac{\mu_{2}^{2}+\mu_{3}^{2}}{2}+\frac{(\mu_{2}^{2}-\mu_{3}^{2})\cos4\varphi_{\vec{q}}}{2}}}\label{EqSoundVelocity}
\end{equation}
and their polarizations are:

\begin{align}
\vec{e}_{+}\left(\varphi\right) & =\frac{1}{\sqrt{2x\left(x+\mu_{2}\cos2\varphi\right)}}\begin{pmatrix}\mu_{2}\cos2\varphi+x\\
\mu_{3}\sin2\varphi
\end{pmatrix}\nonumber \\
\vec{e}_{-}\left(\varphi\right) & =\frac{1}{\sqrt{2x\left(x+\mu_{2}\cos2\varphi\right)}}\begin{pmatrix}-\mu_{3}\sin2\varphi\\
\mu_{2}\cos2\varphi+x
\end{pmatrix}\label{polarizations}
\end{align}
where $\rho$ is the density and $x=\sqrt{\frac{1}{2}(\mu_{2}^{2}+\mu_{3}^{2})+\frac{1}{2}(\mu_{2}^{2}-\mu_{3}^{2})\cos(4\varphi)}$.
In the remainder of the paper, we focus our analysis on the low-energy
phonon mode, $\omega_{\vec{q},-}$, and drop the branch index $\lambda$.
The anisotropy of this mode, which is four-fold symmetric, depends
on the relative strength of the elastic constants, since $\frac{v_{s}\left(0\right)}{v_{s}\left(\pi/4\right)}=\sqrt{\frac{2C_{66}}{C_{11}-C_{12}}}$.
As a result, the phonon anisotropy is stronger in systems close to
a tetragonal-to-orthorhombic lattice instability, since in this case
either $C_{66}\rightarrow0$ (corresponding to a $B_{2g}$ deformation
of the square lattice) or $C_{11}-C_{12}\rightarrow0$ (corresponding
to a $B_{1g}$ deformation of the square lattice). Because the phase
diagrams of the iron pnictides, of some cuprates, and of some heavy
fermions display nematic fluctuations \cite{FernandesNP2014,FradkinARCMP2010}
related to a $B_{1g}$ tetragonal-to-orthorhombic instability, we
focus on the case $\left(C_{11}-C_{12}\right)/2\ll C_{66}$. In this
situation, the sound velocity is minimum at $\varphi=\pi/4$ and maximum
at $\varphi=0$, as shown in Fig.~\ref{FigPhononLatticeDistortionIllustration}b.
The polarization $\mathbf{e}_{\mathbf{q},-}$ of the low-energy phonon
mode described by Eq. (\ref{EqSoundVelocity}) is transversal at the
high-symmetry propagation directions $\varphi=0,\,\frac{\pi}{4},\,\frac{\pi}{2}$,
implying that the electron-phonon matrix element $M_{\vec{q},-}$
vanishes at these directions. This behavior is shown in Fig.~\ref{FigPhononLatticeDistortionIllustration}c.
Note that the anisotropy of $M_{\vec{q},-}$, given by Eq. (\ref{matrix_element}),
arises both from the anisotropy of the polarization and from the anisotropy
of the sound velocity.

\section{Momentum-dependent effective electronic temperature \label{sec_T_momentum}}

\subsection{Boltzmann formalism}

Having established the properties of the coupling between electrons
and phonons, we now discuss how driving the acoustic phonons out of
equilibrium affects the low-energy electronic states. Experimentally,
a non-equilibrium distribution of acoustic phonons, which we denote
by $n_{B}(\omega_{\vec{q}})$, can be generated by ultrafast strain
of interfaces \cite{MatteoN2007,PezerilPRL2011,CavigliaPRL2012,ForstNM2015,Giannetti-AdvPhys-2016}.
A periodic driving of such non-equilibrium phonons is necessary to
establish a steady-state non-thermal phononic distribution, otherwise
the phonons would relax back to equilibrium. These out-of-equilibrium
phonons inelastically scatter electronic quasi-particles between their
momentum eigenstates, resulting in a non-equilibrium electronic distribution
function $n_{\xi_{\vec{p}}}^{F}$. Theoretically, the latter is given
by the solution of the Boltzmann equation $I_{\mathrm{coll}}^{\mathrm{phonon}}[n^{F},n^{B}]=0$,
where $I_{\mathrm{coll}}^{\mathrm{phonon}}$ denotes the phonon collision
integral: 
\begin{widetext}
\begin{equation}
I_{\mathrm{coll}}^{\mathrm{phonon}}[n^{F},n^{B}]=-\sum_{\alpha=\pm}\alpha\int\frac{\mathrm{d}^{2}p'}{(2\pi)^{2}}\delta(\xi_{\vec{p}}-\xi_{\vec{p}'}-\alpha\omega_{\vec{p}-\vec{p}'})|M_{\vec{p}-\vec{p}'}|^{2}\left[n_{\xi_{\vec{p}}}^{F}n_{-\xi_{\vec{p}'}}^{F}n_{-\alpha\omega_{\vec{p}-\vec{p}'}}^{B}+n_{-\xi_{\vec{p}}}^{F}n_{\xi_{\vec{p}'}}^{F}n_{\alpha\omega_{{\vec{p}-\vec{p}'}}}^{B}\right]\,,\label{EqTheFullFledgedCollisionKernel}
\end{equation}
\end{widetext}

\begin{figure*}
\begin{centering}
\includegraphics[width=0.9\linewidth]{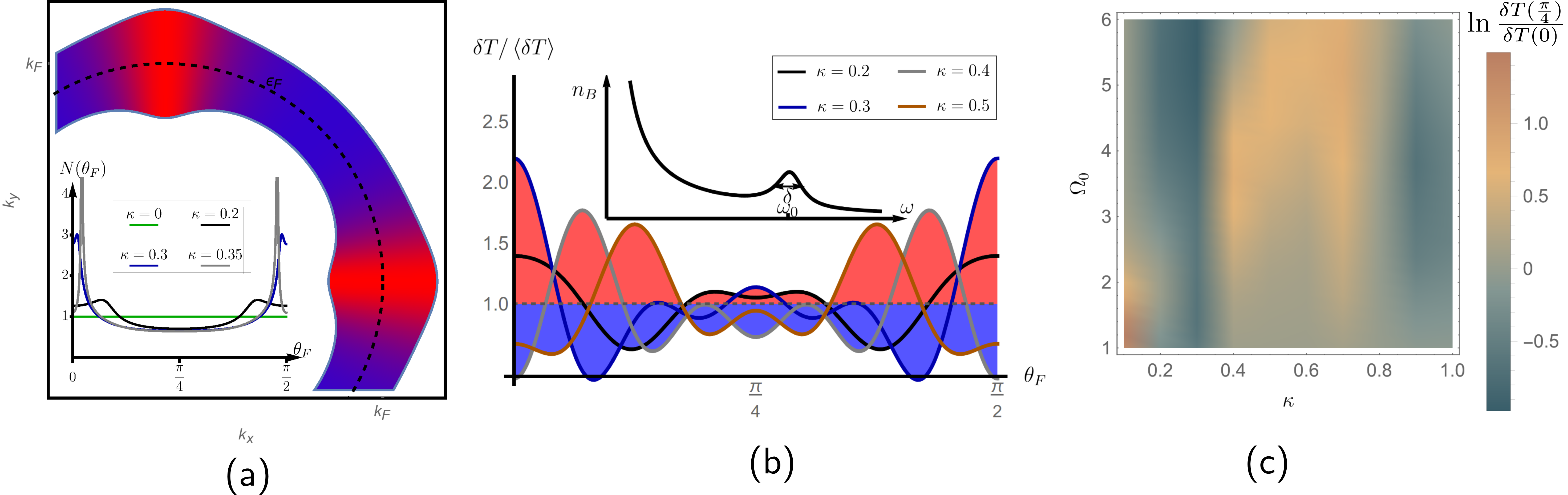} 
\par\end{centering}
\caption{\textbf{Anisotropic redistribution of electronic state and non-uniform
heating of the Fermi surface states.} (a) Due to the geometric constraint
imposed by momentum conservation, Eq.~\eqref{restriction}, the electronic
states capable of absorbing acoustic phonons are not uniformly distributed
around the Fermi surface (black dashed line). States near the red
thick (blue thin) regions have a larger (smaller) phase space for
phonon scattering. The inset shows the density of electronic states
$N\left(\theta_{F}\right)$ available for the scattering of an electron
at a Fermi surface angle $\theta_{F}$ by a phonon of energy $\omega_{0}$,
for different values of the dimensionless parameter $\kappa\equiv\left(\frac{\omega_{0}}{2\varepsilon_{F}}\right)/\left(\frac{v_{s}(0)}{v_{F}}\right)$.
(b) Momentum-dependent temperature profile $\delta T\left(\theta_{F}\right)$,
normalized by the average heating $\left\langle \delta T\right\rangle $,
for different values of $\kappa$. The phonon non-equilibrium distribution
$n_{B}$ is depicted in the inset. We set the ratio $\Omega_{0}\equiv\omega_{0}/2T=5.8$.
Due to the scattering of electrons by non-equilibrium phonons, certain
regions of the Fermi surface are locally hotter (red shade) or colder
(blue shade) than the average. The typical momenta associated with
the hot/cold regions change as function of $\kappa$. (c) Color plot
of the anisotropy of the non-uniform temperature, as defined by the
logarithm of the ratio $\delta T(\pi/4)/\delta T(0)$, as function
of the two dimensionless parameters $\kappa$ and $\Omega_{0}$. Blue
(yellow) denotes dominant heating near $\theta_{F}=0,\,\pi/2$ ($\theta_{F}=\pi/4$).}
\label{FigNumericallyDeterminedTemperatureProfile} 
\end{figure*}

Here, we introduced the convention $n_{-x}^{F}=1-n_{x}^{F}$ and $-n_{-x}^{B}=n_{x}^{B}+1$.
The physical meaning of the Boltzmann equation is clear: in order
for the system to maintain a homogeneous quasi-particle distribution
in the steady-state, any deviation of the phonon distribution function
from the equilibrium Bose-Einstein function $n_{0,\mathbf{p}}^{B}$
must be compensated by a deviation of the electronic distribution
from the Fermi function $n_{0,\mathbf{p}}^{F}$. To focus on the general
properties of the mechanism proposed here, we consider small deviations
from equilibrium, $n_{\mathbf{p}}^{F/B}=n_{0,p}^{F/B}-(\partial_{\xi_{\vec{p}}}n_{0,p}^{F/B})h_{\vec{p}}^{F/B}$,
and linearize the Boltzmann equation in the functions $h_{\mathbf{p}}^{F/B}$.
The resulting integral equation can be conveniently recast as a functional
minimization problem, which allows for a direct determination of the
electronic non-equilibrium distribution function $h^{F}$ for a given
phononic non-equilibrium distribution function $h_{\mathbf{p}}^{B}$
(details in Appendix \eqref{app_boltzmann}). Later in Section \eqref{sec_Discussion}
we discuss the effects of other scattering processes that give additional
contributions to the collision integral, such as impurity scattering
and electron-electron scattering.

\textit{\emph{While the equilibrium electronic distribution $n_{0,\vec{p}}^{F}$
is determined entirely by the chemical potential $\mu$ and the temperature
$T$, the non-equilibrium distribution can be generally parametrized
in terms of a momentum-dependent effective temperature $T+\delta T_{\hat{\vec{p}}}$,
with $\hat{\vec{p}}=\vec{p}/|\vec{p}|$, and an effective chemical
potential $\mu+\delta\mu_{\hat{\vec{p}}}$, yielding $h_{\vec{p}}^{F}=\delta\mu_{\hat{\vec{p}}}+\frac{\xi_{\vec{p}}}{T}\,\delta T_{\hat{\vec{p}}}$.}}
Such a parametrization is particularly useful for the states near
the Fermi level, where the energy scales associated with momenta perpendicular
and parallel to the Fermi surface are very different. In our problem,
because the interaction with a single acoustic phonon cannot modify
the chemical potential locally, $\delta\mu_{\hat{\vec{p}}}$ is zero.
Furthermore, the small energy transfer resulting from the electron-phonon
scattering constrains the low-energy electrons to remain near the
Fermi surface. As a result, the steady-state non-equilibrium electronic
distribution function is completely encoded in the momentum-dependent
effective temperature $\delta T_{\hat{\vec{p}}}$. Note that this
approximation is valid only for a non-current-carrying state; finite
currents would require an additional shift of the momentum states.
We emphasize that $\delta T_{\hat{\vec{p}}}$ is not a thermodynamic
temperature, but a convenient parametrization of the distribution
function. Its effect on physical observables will be studied in further
details in Section \eqref{sec_phase_diagrams}.

The precise momentum dependence of the effective electronic temperature
$\delta T_{\hat{\vec{p}}}$ depends on the type of phonon distribution
function $h_{\mathbf{p}}^{B}$. As expected, in the simple case of
a uniform heating of the phonons, the solution of the Boltzmann equation
gives just a momentum-independent shift of the effective electronic
temperature. \textit{\emph{As we show below, in}} order to induce
anisotropies in $\delta T_{\hat{\vec{p}}}$, it is sufficient to excite
phonons, isotropically, around a well-defined energy $\omega_{0}$, as the geometrical
constraints imposed by momentum and energy conservations, together
with the anisotropy of the sound velocity, cause an anisotropic redistribution
of quasi-particles.

\subsection{Microscopic mechanism for the anisotropic effective temperature}

To illustrate this generic effect, we consider a circular Fermi surface
of radius $p_{F}$, as shown in Fig.~\ref{FigNumericallyDeterminedTemperatureProfile}a.
The Fermi momenta are parametrized by $\mathbf{p}_{F}=p_{F}\left(\cos\theta_{F},\,\sin\theta_{F}\right)$.
Momentum conservation enforces a relationship between the initial
momentum $\theta_{F}$ and the final momentum $\theta'_{F}$, $q=2p_{F}\sin\left(\frac{\theta_{F}-\theta'_{F}}{2}\right)$,
where $q$ is the phonon momentum. Now, the phonon has a well-defined
energy, $q=\omega_{0}/v_{s}\left(\varphi_{\vec{q}}\right)$, and a
well-defined propagation direction $\varphi_{\vec{q}}$, which is
also related to the initial and final momenta by $2\varphi_{\vec{q}}=\pi+\theta_{F}+\theta'_{F}$.
As a result, for a given momentum $\theta_{F}$, the allowed values
for $\theta'_{F}$ are given by the solution of the implicit equation:

\begin{equation}
\kappa=\tilde{v}_{s}\left(\frac{\theta_{F}+\theta'_{F}}{2}\right)\sin\left(\frac{\theta_{F}-\theta'_{F}}{2}\right)\label{restriction}
\end{equation}

Here, for convenience, we introduced the dimensionless parameter $\kappa\equiv\left(\frac{\omega_{0}}{2\varepsilon_{F}}\right)/\left(\frac{v_{s}(0)}{v_{F}}\right)$,
which relates the typical sound velocity $v_{s}\left(0\right)$, the
Fermi velocity $v_{F}$, the excited phonon frequency $\omega_{0}$,
and the Fermi energy $\varepsilon_{F}$. The normalized phonon velocity,
defined as $\tilde{v}_{s}\left(\varphi\right)\equiv v_{s}\left(\varphi\right)/v_{s}\left(0\right)$,
is by definition always smaller than $1$, since the sound velocity
is maximum at $\varphi=0$. The key point of Eq. (\ref{restriction})
is that the parameter $\kappa$ strongly affects the allowed values
for the pair of momenta $\left(\theta_{F},\theta'_{F}\right)$. For
example, when $\kappa\ll1$ the solution of Eq. (\ref{restriction})
requires the initial and final momenta to be very close, $\theta_{F}'\approx\theta_{F}$,
whereas when $\kappa>\frac{v_{s}\left(\pi/4\right)}{v_{s}\left(0\right)}$,
there is no pair of momenta $\left(\theta_{F},\theta'_{F}\right)$
that solves Eq. (\ref{restriction}). Note the key role played by
the anisotropy in the sound velocity $\tilde{v}_{s}\left(\frac{\theta_{F}+\theta'_{F}}{2}\right)$:
without it, the equation would only depend on the relative momentum
$\theta_{F}-\theta'_{F}$.

For intermediate values of $\kappa$, the geometric constraint imposed
by Eq.~\eqref{restriction} implies that the electronic states that
can absorb an acoustic phonon are not equally distributed around the
Fermi surface. To quantify this important property, we use Eq.~\eqref{restriction}
to compute the density of available states, $N\left(\theta_{F}\right)$,
for an electron with momentum $\theta_{F}$ scattered by a phonon
of energy $\omega_{0}$. The derivation of $N\left(\theta_{F}\right)$
is presented in Appendix \eqref{app_constraint}. In Fig.~\ref{FigNumericallyDeterminedTemperatureProfile}a,
we show the behavior of $N\left(\theta_{F}\right)$ projected along
the Fermi surface for $\kappa=0.3$: it is clear that Fermi surface
states with $\theta_{F}=0,\,\pi/2$ are much more efficient in absorbing
the acoustic phonons as compared to the states with $\theta_{F}=\pi/4$.
Consequently, the effective temperature along $\theta_{F}=\pi/4$,
as caused by phonon scattering, will generally be smaller than along
$\theta_{F}=0,\pi/2$. Moreover, the fact that the electron-phonon
matrix vanishes for transferred momentum along $\pi/4$ (see Fig.
\eqref{FigPhononLatticeDistortionIllustration}c) causes an additional
suppression of the effective temperature along $\theta_{F}=\pi/4$.
This is the microscopic origin of the momentum-dependent effective
temperature induced by non-equilibrium acoustic phonons. Note from
the inset of Fig.~\ref{FigNumericallyDeterminedTemperatureProfile}a
that the region of the Fermi surface that is more affected by phonon
scattering changes continuously as function of $\kappa$, becoming
narrower as $\kappa$ approaches the limiting value $\frac{v_{s}\left(\pi/4\right)}{v_{s}\left(0\right)}\approx0.5<1$.
Consequently, the degree of anisotropy in the effective temperature
is controlled by the energy of the excited phonons $\omega_{0}$.
Since $\kappa\sim\mathcal{O}(1)$, the relevant phonon energies are
always much smaller than the Fermi energy, $\omega_{0}/\varepsilon_{F}\sim\mathcal{O}(v_{s}(0)/v_{F})$.

\subsection{Boltzmann result for the anisotropic effective temperature}

The general analysis above is confirmed by explicit solution of the
Boltzmann equation for $\delta T_{\mathbf{p}}$. In the case of phonons
excited near an energy $\omega_{0}$, the non-equilibrium bosonic
distribution function is modeled as (see the inset in Fig.~\ref{FigNumericallyDeterminedTemperatureProfile}b):

\begin{equation}
h_{\vec{q}}^{B}=W_{B}\,\frac{\delta}{\delta^{2}+\bigl(\frac{\omega_{\vec{q}}}{2T}-\Omega_{0}\bigr)^{2}}\,.\label{phonon_hB}
\end{equation}

Here, the parameter $\delta\ll1$ represents the energy width of the
excited phonons, $W_{B}$ is an overall amplitude corresponding to
the number of bosons excited by the external drive, and $\Omega_{0}\equiv\frac{\omega_{0}}{2T}$
. In Fig.~\ref{FigNumericallyDeterminedTemperatureProfile}b, we
show the momentum dependence of the effective temperature profiles
$\delta T(\theta_{F})$ obtained by solving the Boltzmann equation
for a fixed $\Omega_{0}$ as function of the parameter $\kappa$.
For small values of $\kappa$, the anisotropy (as measured by the
ratio $\delta T(\frac{\pi}{4})/\delta T(0)$) is mild, with the states
with momentum $\theta_{F}=\pi/4$ only slightly colder than those
with momenta $\theta_{F}=0,\,\pi/2$. Upon increasing $\kappa$, the
anisotropy is clearly increased \textendash{} in particular, the maximum
anisotropy, for $\kappa\approx0.3$, takes place when the $\theta_{F}=0,\,\pi/2$
states are the hottest states at the Fermi surface. Upon further increasing
$\kappa$, the hottest region of the Fermi surface moves back towards
the diagonal, but the amplitude of the anisotropy decreases. Remarkably,
the changes in $\delta T(\theta_{F})$ as function of $\kappa$ are
nearly insensitive to the value of $\Omega_{0}$, as shown in Fig.~\ref{FigNumericallyDeterminedTemperatureProfile}c.

This behavior of $\delta T(\theta_{F})$ is in qualitative agreement
with the geometrical analysis of Fig.~\ref{FigNumericallyDeterminedTemperatureProfile}a,
which shows that the largest density of electronic states capable
of absorbing a phonon moves towards the Fermi surface region near
$\theta_{F}=0,\,\pi/2$ and then back towards $\theta_{F}=\pi/4$
as $\kappa$ increases. The fact that the amplitude of the anisotropy
is not monotonic can be attributed to the fact that the maximum of
the density $N(\theta_{F})$ becomes not only larger but also narrower
as $\kappa$ increases, while the distribution function~\eqref{phonon_hB}
is sensitive to a window of energies centered around $\omega_{0}$.
Furthermore, the small changes near $\theta_{F}=\pi/4$ as function
of $\kappa$ are a consequence of the numerator of the electron-phonon
matrix element in Eq. \eqref{matrix_element} vanishing for transferred
momentum $\pi/4$.

\section{Impact on competing electronic phases\label{sec_phase_diagrams}}

We have seen that electrons that interact with out-of-equilibrium
acoustic phonons develop an anisotropic non-equilibrium distribution
that can be characterized by a momentum dependent effective electronic
temperature. We now show that this can be used as a tool to control
and manipulate competing electronic states in correlated systems.
The idea is that, by tuning the excited phonon energy $\omega_{0}$
(proportional to the parameter $\kappa$ discussed above), one can
in principle selectively heat certain regions of the Fermi surface.
While applicable to any form of non-isotropic order, this is particularly
relevant for density-wave instabilities, which are generally governed
by the electronic states near the hot spots $\mathbf{k}_{\mathrm{hs}}$
\textendash{} points of the Fermi surface separated by the density-wave
ordering vector $\mathbf{Q}$, $\xi_{\mathbf{k}_{\text{hs}}}=-\xi_{\mathbf{k}_{\text{hs}}+\mathbf{Q}}$.
In this case, an appropriate momentum-dependent effective temperature
profile can be applied to selectively melt the density-wave state
while preserving other homogeneous ordered phases.

\begin{figure*}
\begin{centering}
\includegraphics[width=0.7\linewidth]{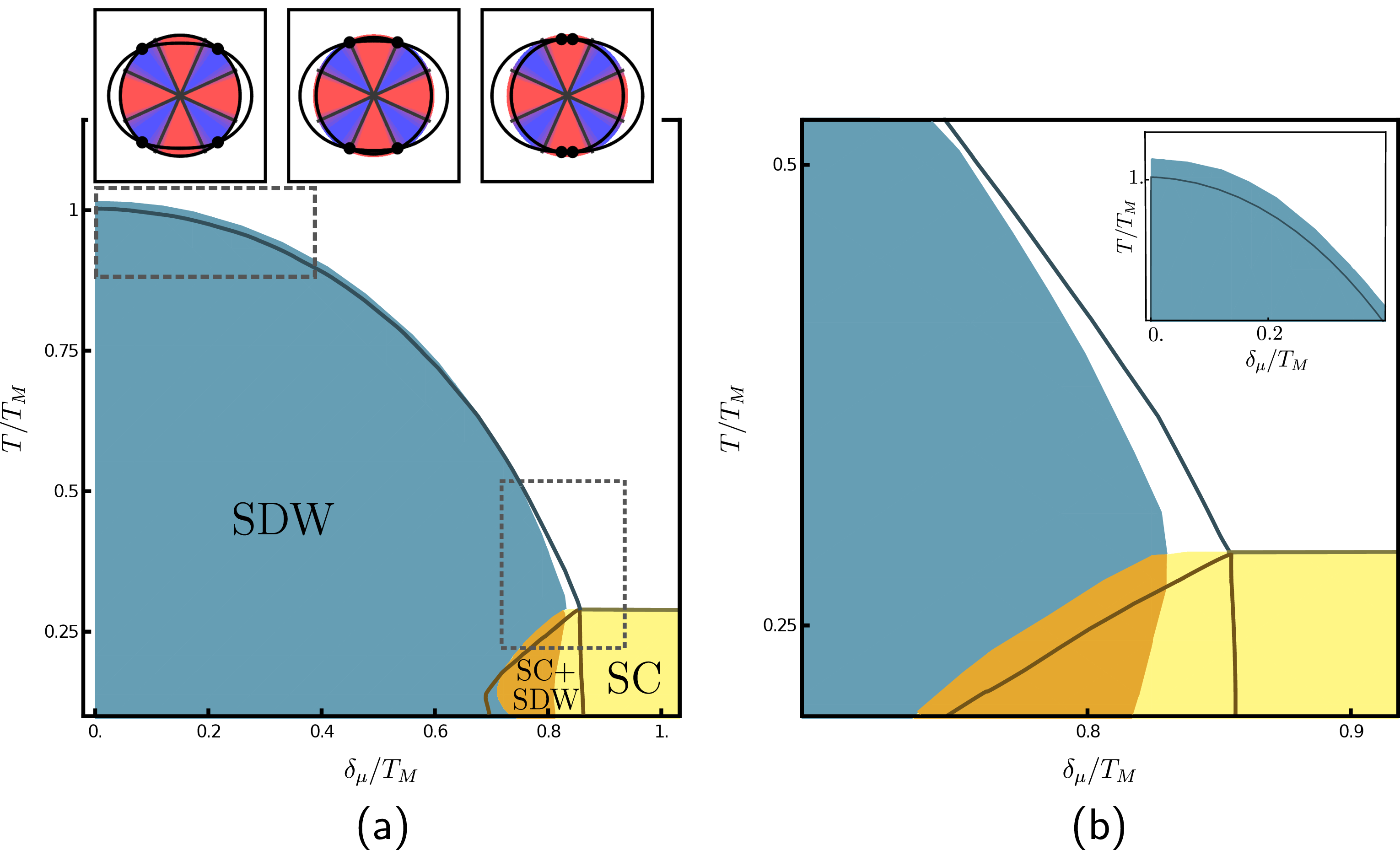} 
\par\end{centering}
\caption{\textbf{Non-equilibrium steady-state phase diagram for competing superconductivity
(SC) and spin-density wave (SDW). }The steady-state non-equilibrium
phase diagram for the two-band model of the iron pnictides is shown
in panel (a), and the highlighted regions are zoomed in panel (b). It contains phases with SDW (cyan), SC (yellow) and coexistent SDW+SC (orange) order.  The
inset shows the superimposed circular hole-like and elliptical electron-like
Fermi pockets, with the latter displaced by the SDW ordering vector
$\mathbf{Q}$. The crossing points are the hot spots $\vec{k}_{\text{hs}}$,
which govern the SDW instability; their positions are controlled by
the parameter $\delta_{\mu}$, which also determines the carrier concentration.
The non-equilibrium phase diagram was obtained for the anisotropic
effective temperature profile $\delta T\left(\theta_{F}\right)$ shown
in Fig. \ref{FigNumericallyDeterminedTemperatureProfile}b (here we set $\kappa=0.27$)
and illustrated in the inset (red and blue denote regions hotter and
colder than the average heating $\left\langle \delta T\right\rangle $,
respectively). For comparison, we also show the equilibrium phase
diagram (solid lines) shifted by the average heating $\left\langle \delta T\right\rangle /T_{M}\approx0.1$.
Here, we set $\delta_{m}/T_{M}=0.9$. All quantities are expressed
in terms of the equilibrium SDW transition temperature $T_{M}$ for
$\delta_{\mu}=0$. }
\label{FigPhononInducedPhaseDiagram} 
\end{figure*}

Although generic, this concept of selective melting can be nicely
demonstrated by an explicit calculation for the case of competing
spin-density wave (SDW) and $s^{+-}$ superconductivity (SC) in the
iron pnictides. We emphasize that our goal here is not to provide
a microscopically calculated phase diagram for a specific iron pnictide
material, but rather to use a transparent low-energy model to demonstrate
the general concept proposed here. In this spirit, an effective low-energy
model widely employed to study the SDW-SC competition in the pnictides
is a two-band model with one circular hole pocket at the center of
the Fe square-lattice Brillouin zone, and one elliptical electron
pocket centered at the SDW ordering vector $\mathbf{Q}$~\cite{FernandesRPP2017}.
The non-interacting Hamiltonian $H_{0}$ then contains two bands that
can be conveniently parametrized as $\xi_{h,\mathbf{k}}=\varepsilon_{0}-\frac{k^{2}}{2m}$
and $\xi_{e,\mathbf{k}+\mathbf{Q}}=-\xi_{h,\mathbf{k}}-\delta_{\mu}-\delta_{m}\cos2\theta$.
The parameters $\delta_{\mu}$ (proportional to the electronic occupation
number) and $\delta_{m}$ (proportional to the ellipticity of the
electron pocket) serve as a measure of the nesting condition between
the two bands.

The two leading instabilities arising from $H_{\mathrm{el}}$ are
the $s^{+-}$ SC instability, characterized by two uniform gaps of
opposite signs in the two bands, and the SDW instability \cite{FernandesRPP2017}.
The corresponding interactions projected onto these two channels are
denoted by $V_{\mathrm{SC}}$ and $V_{\mathrm{SDW}}$, respectively.
While the former is sensitive to all Fermi surface states, the latter
is governed by the hot spots. When $|\delta_{\mu}|<|\delta_{m}|$,
there are four pairs of hot spots located at the Fermi surface angles
multiples of $\theta_{\mathrm{hs}}=\frac{1}{2}\arccos(-\delta_{\mu}/\delta_{m})$,
whereas when $\left|\delta_{\mu}\right|>\left|\delta_{m}\right|$,
there are no hot spots. Thus, for a fixed $\delta_{m}$, increasing
$\left|\delta_{\mu}\right|$ makes nesting poorer, which suppresses
the SDW instability. The $\left(\delta_{\mu},T\right)$ equilibrium
phase diagram\textit{ }of this model is shown by the solid lines in
Fig.~\ref{FigPhononInducedPhaseDiagram} for fixed $\delta_{m}$.
In this plot, the transition lines have been shifted to mimic a non-zero
average uniform heating, as explained below.

To obtain the phase diagram of the non-equilibrium steady state in
which the electrons interact with a distribution of non-equilibrium
acoustic phonons, we use the Keldysh formalism. The main result, derived
in Appendix \eqref{app_phase_diagram}, can be explained in terms
of the self-consistent gap equations governing the SC and SDW instabilities.
Let us denote the corresponding order parameters by $\Delta$ and
$M$. In the Keldysh language, they correspond to classical source
fields. The linearized self-consistent equations for $M$ and $\Delta$,
derived using the Keldysh approach, are given by:

\begin{align}
M & =2V_{\mathrm{SDW}}\int\frac{\mathrm{d}^{2}p}{\left(2\pi\right)^{2}}\left[\frac{f\left(\xi_{h,\mathbf{p}}\right)-f\left(\xi_{e,\mathbf{p+Q}}\right)}{\xi_{h,\mathbf{p}}-\xi_{e,\mathbf{p+Q}}}\right]M\nonumber \\
\Delta & =V_{\mathrm{SC}}\int\frac{\mathrm{d}^{2}p}{\left(2\pi\right)^{2}}\left[\frac{f\left(\xi_{h,\mathbf{p}}\right)}{\xi_{h,\mathbf{p}}}+\frac{f\left(\xi_{e,\mathbf{p+Q}}\right)}{\xi_{e,\mathbf{p+Q}}}\right]\Delta\label{eq_gaps}
\end{align}

The key point is that the function $f\left(\xi_{\mathbf{p}}\right)$,
which relates the advanced, retarded, and Keldysh components of the
electronic Green's function, $G^{K}=f\left(G^{R}-G^{A}\right)$, is
given in terms of the non-equilibrium electronic distribution function
$n_{\xi_{\vec{p}}}^{F}$ by $f\left(\xi_{\mathbf{p}}\right)=1-2n_{\xi_{\vec{p}}}^{F}$.
If the system was in equilibrium, the function $f$ would acquire
the familiar form $f\left(\xi\right)=\tanh\left(\frac{\beta\xi}{2}\right)$,
and Eqs. \eqref{eq_gaps} would reduce to the standard equilibrium
gap equations. Out of equilibrium, however, $f\left(\xi_{\mathbf{p}}\right)$
contains information about the non-thermal distribution of electrons,
which in turn is parametrized in terms of the effective momentum-dependent
temperature $T+\delta T_{\hat{\vec{p}}}$. Thus, mathematically, the
Keldysh formalism reveals that the effect of the coupling between
electrons and non-equilibrium acoustic phonons in this problem can
be cast as self-consistent gap equations that have the same form as
the equilibrium gap equations, but with the thermodynamic temperature
$T$ replaced by the effective temperature $T+\delta T_{\hat{\vec{p}}}$,
where $T$ is the temperature of a bath to which both the acoustic
phonons and the electrons are coupled (such a bath could be due to
optical phonons or other degrees of freedom in the system). This justifies
associating $\delta T_{\hat{\vec{p}}}$ to an effective electronic
temperature. Note that this microscopic approach recovers the phenomenological
one introduced by Eliashberg and others in Refs. \cite{Eliashberg,EliashbergInLangenbergLargkin1986},
which correctly predicted the enhancement of $T_{c}$ in SC thin films
irradiated by microwaves.

We can now use these gap equations to calculate the SC-SDW phase diagram
for electrons subject to a momentum dependent effective temperature
$\delta T\left(\theta_{F}\right)$. This should be understood, in
the same spirit as Refs. \cite{Eliashberg,EliashbergInLangenbergLargkin1986},
as a steady-state non-equilibrium phase diagram, since without driving, at long enough
times, the system will relax back to equilibrium. In this regard,
the transition temperatures correspond to the temperatures of the
bath discussed in the paragraph above. Because the linearized gap
equations \eqref{eq_gaps} can only give the leading instability of
the system, to study the competition between SC and SDW we must also
included higher order terms of the non-linear gap equations, as explained
in Appendix \eqref{app_phase_diagram}.

In what follows, we use the profile $\delta T\left(\theta_{F}\right)/\left\langle \delta T\right\rangle $
calculated in the previous section for $\kappa=0.27$ and $\Omega_{0}=5.8$
, corresponding to dominant heating around $\theta_{F}=0,\,\pi/2$
(see Fig.~\ref{FigNumericallyDeterminedTemperatureProfile}b). The
steady-state phase diagram is shown by the shaded regions in Fig.~\ref{FigPhononInducedPhaseDiagram}.
To make the comparison with the equilibrium phase diagram more meaningful,
and to highlight the impact of non-uniform heating, the equilibrium
temperature was shifted by the average heating $\left\langle \delta T\right\rangle \equiv\int\frac{d\theta_{F}}{2\pi}\,\delta T\left(\theta_{F}\right)$,
which depends on the external drive.

On the overdoped side of the phase diagram ($\delta_{\mu}/T_{M}\gtrsim0.75$),
where only SC is present, we find that the SC $T_{c}$ is unaffected
by the non-uniform heating. This is because the $s^{+-}$ SC instability
has equal contributions from all electronic states at the Fermi surface,
since the gap function is uniform. As a result, $T_{c}$ is sensitive
only to the average temperature around the Fermi surface, $\left\langle \delta T\right\rangle $.
In contrast, on the very underdoped side of the phase diagram (near
$\delta_{\mu}=0$), where only SDW is present, we observe that the
SDW transition temperature $T_{M}$ increases for non-uniform heating,
because the SDW instability is dominated by the hot spots of the Fermi
surface. For these doping levels, as shown in the inset of Fig.~\ref{FigPhononInducedPhaseDiagram}a,
the hot spots are located near $\theta_{F}=\pi/4$, where the local
effective temperature $\delta T(\frac{\pi}{4})$ oscillates weakly
around the average $\left\langle \delta T\right\rangle $. Electrons near the hot spots experience less heating than the average. 

As $\delta_{\mu}$ (i.e. doping) increases, the hot spots move towards
the $\theta_{F}=\pi/2$ direction; this behavior is also seen experimentally
in certain pnictides~\cite{LiuNP2010}. Because the local temperature
in this Fermi surface region is larger than the average one, $T_{M}$
decreases. More interestingly, as highlighted in Fig.~\ref{FigPhononInducedPhaseDiagram}b,
superconductivity is favored and $T_{c}$ systematically increases
near the optimal doping regime $\delta_{\mu}/T_{M}\approx0.75$, as
compared to its equilibrium value. Because the SC order is unaffected
by the non-uniform heating, this enhancement of $T_{c}$ is a direct
consequence of the suppression of the SDW transition $T_{M}$. This clearly illustrates 
that a momentum-dependent effective electronic temperature is able
to shift the balance between competing orders.

\section{Discussion and conclusions \label{sec_Discussion}}

In this paper we explored a novel framework to manipulate and control
correlated electronic systems out of equilibrium. In particular, we
demonstrated that non-equilibrium acoustic phonons excited around
a well-defined energy lead to a non-uniform redistribution of electronic
states around the Fermi surface. This manifests as a momentum-dependent
effective temperature characterizing the non-equilibrium distribution
function. Our theoretical result is robust, as it stems from geometric
constraints imposed by energy and momentum conservation, together
with the intrinsic anisotropy of the sound velocity. It reveals a
hitherto unexplored path to manipulate correlated states via pump-and-probe
experiments, which have so far been mostly focusing on the coherent
excitation of optical phonons. In contrast to the latter, heating
effects are expected to be much weaker in the case of acoustic phonons,
since the energies involved are significantly smaller.

The application of this interesting idea to realistic systems requires
several conditions to be satisfied. First, it is necessary to create
a non-thermal distribution of acoustic phonons. Recent experimental
advances in the ultrafast strain manipulation of interfaces and heterostructures,
including the production of shock waves, provide a promising route
forward \cite{MatteoN2007,PezerilPRL2011,CavigliaPRL2012,ForstNM2015,Giannetti-AdvPhys-2016}.
While in this paper we focused on the excitation of phonons with a
well-defined energy, similar results will hold for phonons with
a well-defined momentum propagation. Second, this non-thermal phonon
population has to be periodically driven, since the phonons themselves
thermalize, usually in the time scale of several hundrets of picoseconds~\cite{KlettArxiv2017}.
Third, it is necessary to ensure that other scattering processes do
not relax the electronic momenta too fast, washing away the anisotropic
effective temperature induced by the coupling to the non-equilibrium
acoustic phonons.

This is a crucial condition that deserves further consideration. The
usual scattering processes considered other than scattering by phonons
(or by other collective bosonic modes) are impurity scattering and
electron-electron scattering. Their effect on the electronic distribution
function can be obtained by including the corresponding collision
integrals in the Boltzmann equation. While a detailed solution of
the full Boltzmann equation is beyond the scope of this work, there
are important qualitative points that can be made (details in Appendix
\eqref{app_impurity}). In the case of impurity scattering, as long
as the energy of the excited phonons is comparable to the thermodynamic
temperature ($\Omega_{0}=\omega_0/(2T)\gtrsim1$ in Fig. \ref{FigNumericallyDeterminedTemperatureProfile}c),
the anisotropy of the effective electronic temperature generated by
the coupling to the acoustic phonons should persist even when the
electron-impurity matrix element is of the same order as the electron-phonon
matrix element. Importantly, both matrix elements can in principle
be controlled: while the former is proportional to the concentration
of impurities (which is suppressed by annealing), the latter is inversely
proportional to the sound velocity (which is enhanced near a structural
phase transition).

As for the case of electron-electron scattering, it is essential to
distinguish two different processes, namely, energy relaxation and
momentum relaxation. The former causes the electronic subsystem to
thermalize, and is usually assumed to be very fast (in the femtosecond
time scale) in phenomenological models widely employed to describe
the relaxation of materials taken out of equilibrium, such as the
two-temperature model. Nevertheless, such an assumption of a very
fast electron thermalization may not hold for several systems. Theoretically,
previous works studying the solution of the Boltzmann equation with
both electron-phonon and electron-electron scattering \cite{Groeneveld95,Kabanov-PRB-2008}
revealed that, for a wide range of parameters, electron-electron scattering
is actually unable to bring the electrons to a thermal distribution
in the time scales of the electron-phonon relaxation time. Experimental
signatures of this effect were observed early on for several simple
metals \cite{Groeneveld95}. More recently, it has been argued that
a similar effect may take place in cuprates \cite{Giannetti-AdvPhys-2016}.
Importantly, these works show that the electron-electron energy relaxation
time can become comparable or even longer than the electron-phonon
relaxation time, depending on materials properties and experimental
conditions. For the framework that we propose here, because the relevant
electronic states remain near the Fermi energy due to the small energy
transferred by the acoustic phonons, the electron-electron energy
relaxation is expected to be longer due to Pauli's principle. Regardless
of its time scale, it is important to note that this energy relaxation
process does not wash away the anisotropy of the effective electronic
temperature, as it does not promote momentum relaxation. In fact,
it actually helps establishing an effective electronic temperature,
as it brings the electrons to a nearly thermal distribution, which
justifies the expansion of the non-equilibrium electronic distribution
function around a thermal distribution (as we assumed in Sec. \ref{sec_T_momentum}A).
The process that is harmful to the anisotropy of the effective temperature
is the electron-electron momentum relaxation, which generally takes
place on longer time scales than the energy relaxation. In two-dimensional
systems, the difference between the two relaxation times is expected
to be even larger, due to the dominance of scattering processes involving
zero momentum transfer \cite{PRB69121102}.

Among several possible applications of the anisotropic effective electronic
temperature generated by non-equilibrium acoustic phonons, we showed
here that it can be employed to selectively melt competing electronic
states, particularly in the case of unconventional superconductivity
competing with a density-wave type of order. The potential enhancement
of $T_{c}$ by non-equilibrium acoustic phonons complements previous
approaches in which SC is enhanced by optical phonons or microwave
radiation. These results also open a broad set of questions that deserve
further investigation. An interesting issue is the impact of non-equilibrium
acoustic phonons on electronic orders that directly couple to the
lattice, such as nematic orders and charge-density waves. Furthermore,
nodal superconducting states are ideal candidates to be manipulated
by a momentum-dependent effective temperature, since the nodal quasi-particles,
which determine the low-energy excitation spectrum, can experience
a different local effective temperature than the average. Finally,
while here we focused on long-range order, a non-uniform effective
electronic temperature should also impact the low-energy charge and
magnetic fluctuation spectra, which influence the pairing state. 
\begin{acknowledgments}
We acknowledge fruitful discussions with A. Chubukov, C. Giannetti,
J. Schmalian, and I. Vishik. M.S. and R.M.F. were supported by the
U.S. Department of Energy, Office of Science, Basic Energy Sciences,
under Award number DE-SC0012336. P.P.O. acknowledges support from
Iowa State University Startup Funds. The work of A.L. was financially
supported by the NSF Grants No. DMR-1606517 and DMR-1653661. Support
for this research at the University of Wisconsin-Madison was provided
by the Office of the Vice Chancellor for Research and Graduate Education
with funding from the Wisconsin Alumni Research Foundation. 
\end{acknowledgments}

\appendix
\bibliographystyle{apsrev4-1}
\bibliography{bibliographyMod}

\section{Solution of the Boltzmann equation \label{app_boltzmann}}

As discussed in the main text, the Boltzmann equation is given by: 
\begin{widetext}
\begin{equation}
I_{\mathrm{coll}}^{\mathrm{phonon}}[n^{F},n^{B}]=-\sum_{\alpha=\pm}\alpha\int\frac{\mathrm{d}^{2}p'}{(2\pi)^{2}}\delta(\xi_{\vec{p}}-\xi_{\vec{p}'}-\alpha\omega_{\vec{p}-\vec{p}'})|M_{\vec{p}-\vec{p}'}|^{2}\left[n_{\xi_{\vec{p}}}^{F}n_{-\xi_{\vec{p}'}}^{F}n_{-\alpha\omega_{\vec{p}-\vec{p}'}}^{B}+n_{-\xi_{\vec{p}}}^{F}n_{\xi_{\vec{p}'}}^{F}n_{\alpha\omega_{{\vec{p}-\vec{p}'}}}^{B}\right]=0\label{1}
\end{equation}
\begin{widetext}
Linearization of the kernel leads to

\begin{align}
\left[n_{\xi_{\vec{p}}}^{F}n_{-\xi_{\vec{p}'}}^{F}n_{-\alpha\omega_{\vec{p}-\vec{p}'}}^{B}+n_{-\xi_{\vec{p}}}^{F}n_{\xi_{\vec{p}'}}^{F}n_{\alpha\omega_{\vec{p}-\vec{p}'}}^{B}\right]\approx-\frac{1}{8}\frac{h_{\xi_{\vec{p}}}^{F}-h_{\xi_{\vec{p}'}}^{F}-\alpha h_{\omega_{\vec{p}-\vec{p}'}}^{B}}{\cosh\left[\frac{\beta\xi_{\vec{p}}}{2}\right]\cosh\left[\frac{\beta\xi_{\vec{p}'}}{2}\right]\sinh\left[\frac{\beta\alpha\omega_{\vec{p}-\vec{p}'}}{2}\right]}
\end{align}
where we used the fact that $h_{\alpha\omega_{q}}^{B}=\alpha h_{\omega_{q}}^{B}$.
Our goal is to find the fermionic distribution $h^{F}$ that solves
the Boltzmann equation for a given phononic distribution $h^{B}$.
It is convenient to work with a functional $\mathcal{F}\left[h^{F}\right]$
whose minimization with respect to $h^{F}$ gives the Boltzmann equation.
We find: 
\begin{multline}
\mathcal{F}\left[h^{F}\right]=\int\frac{\mathrm{d}^{2}p}{(2\pi)^{2}}\int\frac{\mathrm{d}^{2}p'}{(2\pi)^{2}}\sum_{\alpha=\pm}\alpha\delta(\xi_{\vec{p}}-\xi_{\vec{p}'}-\alpha\omega_{\vec{p}-\vec{p}'})\frac{|M_{\vec{p}-\vec{p}'}|^{2}}{16}\frac{\left(h_{\xi_{\vec{p}}}^{F}-h_{\xi_{\vec{p}'}}^{F}\right)^{2}-2\alpha h_{{\vec{p}-\vec{p}'}}^{B}h_{\xi_{\vec{p}}}^{F}+2\alpha h_{{\vec{p}'-\vec{p}}}^{B}h_{\xi_{\vec{p}'}}^{F}}{\cosh\left[\frac{\xi_{\vec{p}}}{2T}\right]\cosh\left[\frac{\xi_{\vec{p}'}}{2T}\right]\sinh\left[\frac{\xi_{\vec{p}}-\xi_{\vec{p}'}}{2T}\right]}\label{EqBarFunctional}
\end{multline}

To proceed, we note that, since the energy of the excited acoustic
phonons is much smaller than the Fermi energy, we can linearize the
electronic dispersion in the vicinity of the Fermi surface. Then,
it is convenient to split the momentum into components perpendicular
and parallel to the Fermi surface (FS), yielding $\int\mathrm{d}^{2}p/(2\pi)^{2}=\int\mathrm{d}\theta/(2\pi)N_{F}(\theta)\int\mathrm{d}\xi_{p}$.
As a result, the electronic states close to the Fermi level, the phonon
dispersion, and the electron-phonon-matrix element Eq.~\eqref{matrix_element}
depend only on the transferred momenta longitudinal to the Fermi surface
$\theta$ and $\theta'$. For simplicity, hereafter we will keep the
notation $\omega_{q}=\omega_{\vec{p}_{F}(\theta)-\vec{p}_{F}(\theta')}$.
Finally, introducing the parametetrization $h_{\vec{p}}^{F}=\delta\mu_{\hat{\vec{p}}}+\frac{\xi_{\vec{p}}}{T}\,\delta T_{\hat{\vec{p}}}$,
we can evaluate the energy integration in the functional Eq.~\eqref{EqBarFunctional},
obtaining a functional that depends only on the longitudinal momenta:
\begin{multline}
\mathcal{F}\left[h^{F}\right]=\iint\frac{\mathrm{d}\theta\mathrm{d}\theta'}{(2\pi)^{2}}N_{F}(\theta)N_{F}(\theta')\frac{\beta\omega_{q}|M_{\vec{q}}|^{2}}{4\sinh^{2}(\frac{\beta\omega_{q}}{2})}\left\{ \left(\frac{\beta\omega_{q}}{2}\right)^{2}\left[\left(\delta T_{\hat{\vec{p}}}+\delta T_{\hat{\vec{p}}'}\right)\right]^{2}\right.\\
\left.+\frac{1}{3}\left[\pi^{2}+\left(\frac{\beta\omega_{q}}{2}\right)^{2}\right]\left(\delta T_{\hat{\vec{p}}}-\delta T_{\hat{\vec{p}}'}\right)^{2}-\beta\omega_{q}h_{\vec{q}}^{B}\left[\delta T_{\hat{\vec{p}}}+\delta T_{\hat{\vec{p}}'}\right]\right\} \label{EqFunctionalAbsorbtionEmmisionSummationn}
\end{multline}

Because the effective chemical potential $\delta\mu_{\hat{\vec{p}}}$
does not appear in the functional, it follows that a non-equilibrium
distribution of acoustic phonons cannot change the chemical potential.
Note that while small momentum scattering is suppressed by the electron-phonon
matrix element Eq~\eqref{matrix_element}, this effect is compensated
by the amount of available thermal states to scatter, which is proportional
to $1/\sinh^{2}(\beta\omega_{q}/2)$.

To minimize the functional, it is convenient to use the Fourier representation
of $\delta T\left(\theta\right)$: \begin{subequations} 
\begin{align}
\delta T(\theta)-\delta T(\theta') & =-2\sum_{n}\delta T_{n}\sin\left(n\frac{\theta-\theta'}{2}\right)\sin\left(n\frac{\theta+\theta'}{2}\right)\label{EqExpressingTheFermionicdependencies}\\
\delta T(\theta)+\delta T(\theta') & =2\sum_{n}\delta T_{n}\cos\left(n\frac{\theta-\theta'}{2}\right)\cos\left(n\frac{\theta+\theta'}{2}\right)
\end{align}
\end{subequations} The Fourier components $\delta T_{n}$ can then
be found by solving the matrix equation $\sum_{m}K_{nm}\delta T_{m}=D_{n}$,
with: 
\begin{align}
D_{n}=-\iint\frac{\mathrm{d}\theta\mathrm{d}\theta'}{(2\pi)^{2}}N_{F}(\theta)N_{F}(\theta')\frac{\beta\omega_{q}|M_{\vec{q}}|^{2}}{4\sinh^{2}(\frac{\beta\omega_{q}}{2})}2\beta\omega_{q}h^{B}(\omega_{q})\cos\left(n\frac{\theta-\theta'}{2}\right)\cos\left(n\frac{\theta+\theta'}{2}\right)\label{EqDrivingExplicit}
\end{align}
and 
\begin{multline}
K_{nm}=\iint\frac{\mathrm{d}\theta\mathrm{d}\theta'}{(2\pi)^{2}}N_{F}(\theta)N_{F}(\theta')\frac{\beta\omega_{q}|M_{\vec{q}}|^{2}}{4\sinh^{2}(\frac{\beta\omega_{q}}{2})}\\
\left\{ \left(\frac{\beta\omega_{q}}{2}\right)^{2}4\cos\left(n\frac{\theta-\theta'}{2}\right)\cos\left(n\frac{\theta+\theta'}{2}\right)\cos\left(m\frac{\theta-\theta'}{2}\right)\cos\left(m\frac{\theta+\theta'}{2}\right)\right.\\
\left.+\frac{1}{3}\left[\pi^{2}+\left(\frac{\beta\omega_{q}}{2}\right)^{2}\right]4\sin\left(n\frac{\theta-\theta'}{2}\right)\sin\left(n\frac{\theta+\theta'}{2}\right)\sin\left(m\frac{\theta-\theta'}{2}\right)\sin\left(m\frac{\theta+\theta'}{2}\right)\right\} \label{EqKernelExplicit}
\end{multline}
\end{widetext}

\end{widetext}

In the main text, we numerically solved these equations using the
following expressions for the phononic distribution function and matrix
element: 
\begin{align}
h^{B}(\omega_{q}) & =W_{B}\,\frac{\delta}{\delta^{2}+\left[\frac{\omega_{q}}{2T}-\Omega_{0}\right]^{2}}\,\\
\omega_{q}|M_{\vec{q}}|^{2} & =m_{0}(\theta,\theta')q^{2}=\tilde{m}_{0}(\theta,\theta')\sin\left(\frac{\theta-\theta'}{2}\right)
\end{align}

Note that the behavior of the overall function $m_{0}(\theta,\theta')$ is given by Fig.~\ref{FigPhononLatticeDistortionIllustration} with $2\varphi=\theta+\theta'$ and that for $\varphi\neq 0,\pi/4,\pi/2$ drops from the equations, since
it appears as a weight in both $K_{nm}$ and $D_{n}$. The phonon
energy is given in terms of the dimensionless parameters and the function
$\tilde{v}_{s}$ defined in the main text: 
\begin{equation}
\frac{1}{2}\beta\omega_{q}=\frac{\Omega_{0}}{\kappa}\,\tilde{v}_{s}\left(\frac{\theta+\theta'}{2}\right)\sin\left(\frac{\theta-\theta'}{2}\right)
\end{equation}

Consequently, the functional depends only on three parameters, $\Omega_{0}$,
$\delta$, and $\kappa$. Prefactors appearing in $h^{B}$, in the
matrix element $|M_{\vec{q}}|^{2}$, and in the density of states
can be conveniently absorbed into the average heating $\left\langle \delta T\right\rangle $.

\section{Analysis of the geometric constraint \label{app_constraint}}

In the main text, we derived the geometric constraint on the initial
and final electronic momenta $\theta_{F}$, $\theta_{F}'$ due to
the energy-momentum conservation associated with electron-phonon scattering:

\begin{equation}
\kappa=\tilde{v}_{s}\left(\frac{\theta_{F}+\theta'_{F}}{2}\right)\sin\left(\frac{\theta_{F}-\theta'_{F}}{2}\right)\label{restriction-1}
\end{equation}

An important quantity is the density of available scattering states
$\theta'_{F}$ for a given momentum $\theta_{F}$, which we denoted
by $N(\theta_{F})$. For a given angle $\theta_{0}$ in the first
quadrant, there are at least two angles $\theta_{1}$ and $\theta_{2}$
also in the first quadrant that satisfy Eq.~\eqref{restriction-1},
with $\theta_{1}>\theta_{0}>\theta_{2}$. In order to determine $N(\theta_{F})$,
let us introduce two ``rotated'' variables $\delta\theta=(\theta_{1}-\theta_{2})/2$
and $\varphi=(\theta_{1}+\theta_{2})/2$, such that condition~\eqref{restriction-1}
becomes $\kappa=\tilde{v}_{s}(\varphi)\sin(\delta\theta)$. Note that
$\varphi$ is, up to a translation by $\pi/2$, the angle corresponding
to the phonon propagation direction. In terms of these variables,
it is straightforward to obtain the solution to Eq. \eqref{restriction-1},
$\delta\theta(\varphi)=\mathrm{arcsin}(\kappa/\tilde{v}_{s}(\varphi))$.
Thus, for a given phonon direction $\varphi$, the two electronic
scattering angles are given by $\theta_{1/2}=\Phi_{\pm}(\varphi)$,
with $\Phi_{\pm}(\varphi)=\varphi\pm\delta\theta(\varphi)$.

It is now straightforward to count the number of all possible electronic
scattering pairs by integrating over all available phonon directions:
\begin{equation}
N(\theta)=\frac{1}{2}\sum_{\alpha=\pm1}\int\mathrm{d}\varphi\,\delta\left[\theta-\Phi_{\alpha}(\varphi)\right].
\end{equation}
yielding: 
\begin{equation}
N(\theta)=\frac{1}{2}\sum_{\alpha=\pm1}\frac{1}{1+\alpha\left(\frac{\partial\delta\theta}{\partial\varphi}\right)_{\varphi=\Phi_{\alpha}^{-1}(\theta)}}
\end{equation}

Alternatively, we can also express the density of scattering states
as function of the phonon direction $\varphi$. In this case, 
\begin{align}
N\left(\varphi\right) & =\frac{1}{2}\left[\left(\frac{\partial\phi(\varphi)}{\partial\varphi_{q}}\right)^{-1}+\left(\frac{\partial\phi'(\varphi)}{\partial\varphi_{q}}\right)^{-1}\right]\nonumber \\
 & =\frac{1}{1-\frac{\kappa^{2}}{\tilde{v}_{s}^{2}-\kappa^{2}}\left(\frac{\tilde{v}_{s}'}{\tilde{v}_{s}}\right)^{2}}.\label{eq_N_phi}
\end{align}

In Fig. 2 of the main text, we plotted $N\left(\theta\right)$ for
different values of $\kappa$. Note that for $\kappa>\mathop{\mathrm{Min}}\limits _{\varphi\in[0,\pi/2]}\sqrt{\frac{v_{s}^{4}}{v_{s}^{2}+(v_{s}')^{2}}}$
the expressions above do not apply and the density needs to be redefined,
since for a given angle $\varphi$, four different pairs $(\theta_{F},\theta_{F}')$
exist. In our case this happens for about $\kappa\approx0.365$.

\begin{figure}[ht]
\includegraphics[width=0.95\columnwidth]{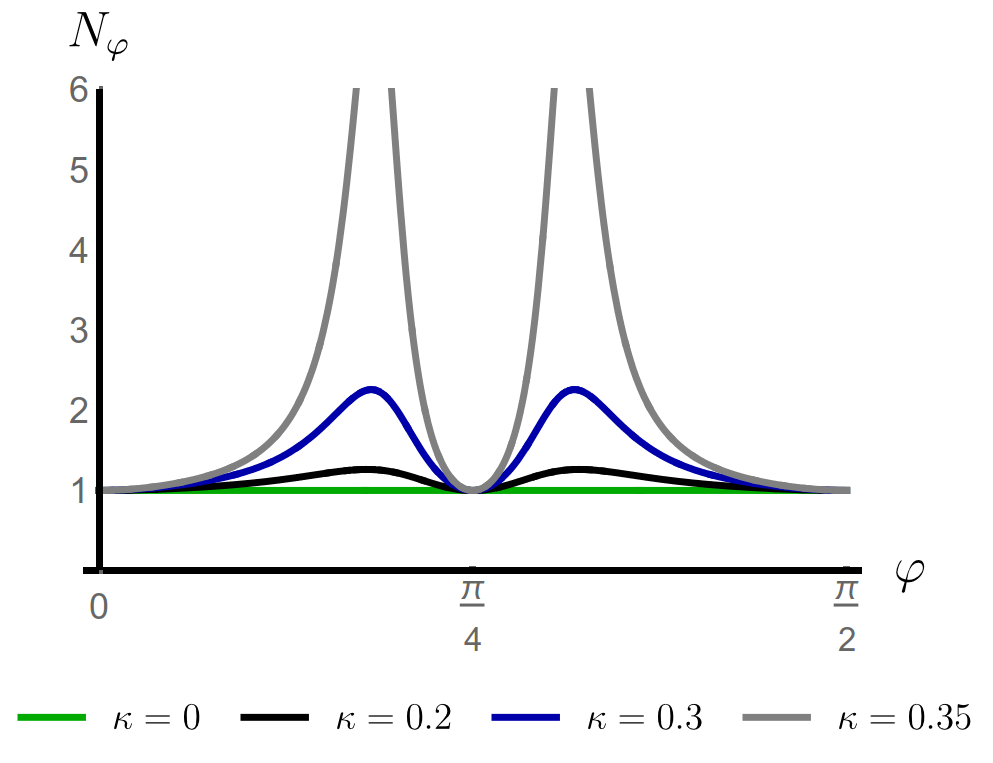} \protect\caption{The scattering density as a function of the phonon direction $\varphi$,
as given by Eq.~\eqref{eq_N_phi}.}
\label{FigScatteringDesnityForPhoninicAngles} 
\end{figure}

In Fig.~\ref{FigScatteringDesnityForPhoninicAngles}, we plot the
density of available scattering solutions as function of the phonon
direction $\varphi$, $N\left(\varphi\right)$. Clearly, for all values
of $\kappa$, the main contribution comes from phonons with $\varphi\approx\varphi_{0}$,
where $\varphi_{0}$ is the angle for which $\sqrt{\frac{v_{s}^{4}}{v_{s}^{2}+(v_{s}')^{2}}}$
is minimal. In our case $\varphi_{0}\approx0.19\pi$. This observation
also explains why the fact that the matrix element vanishes at $\varphi=0,\,\pi/4$
does not cancel the effect, since the dominant phonon directions responsible
for the anisotropic heating do not correspond to the high symmetry
directions $\varphi=0$, $\pi/4$ or $\pi/2$.

\section{Steady-state phase diagram \label{app_phase_diagram}}

The interacting Hamiltonian $H_{\mathrm{int}}$ of the two-band model
contains five different types of inter-pocket and intra-pocket interactions.
To study the competition between SC and SDW, these interactions are
projected in the leading electronic instabilities \cite{FernandesRPP2017}.
The complete phase diagram for the two-band model discussed in the
main text can be obtained by solving the coupled non-linear self-consistent
gap equations for the $s^{+-}$ SC gap, $\Delta=-V_{\mathrm{SC}}\sum_{\vec{k}}\left\langle c_{h,-\vec{k}\downarrow}c_{h,\vec{k}\uparrow}\right\rangle =V_{\mathrm{SC}}\sum_{\vec{k}}\left\langle c_{e,-\vec{k}\downarrow}c_{e,\vec{k}\uparrow}\right\rangle $,
and the SDW order parameter $M=-V_{\mathrm{SDW}}\sum_{\vec{k},\sigma}\left\langle c_{h,\vec{k}\sigma}^{\dagger}\sigma c_{e,\vec{k}\sigma}\right\rangle $.
Here, for convenience, all momenta are measured with respect to the
center of the respective Fermi pocket: $\vec{K}=0$ for the hole pocket
and $\vec{K}=\vec{Q}$ for the electron pocket. The equilibrium phase
diagram of this model was previously calculated in Refs. \cite{FernandesSchmalianPRB2010,VorontsovVavilovPRB2010}.
Here, because we are interested in the normal-state instabilities
of the system, we do not consider the full non-linear gap equations,
but instead expand to cubic order in the order parameters. The reason
why we need to go to cubic order instead of simply linear order is
to capture the effects of the competition between SDW and SC. The
coefficients of the gap equations can be obtained directly from the
Ginzburg-Landau energy functional:

\begin{equation}
F=\frac{a_{M}}{2}M^{2}+\frac{u_{M}}{4}M^{4}+\frac{a_{\Delta}}{2}\Delta^{2}+\frac{u_{\Delta}}{4}\Delta^{4}+\frac{\gamma}{2}M^{2}\Delta^{2}\,.\label{EqGinzburgLandauForm}
\end{equation}

Previously, these coefficients were computed for the two-band microscopic
model in the equilibrium case \cite{FernandesSchmalianPRB2010,VorontsovVavilovPRB2010}.
Our goal here is to show that, in the non-equilibrium steady-state
case, the coefficients have the same functional form, but with the
Fermi-Dirac equilibrium distribution function replaced by the non-equilibrium
distribution function that results from the Boltzmann equation \textendash{}
a procedure widely employed phenomenologically in the literature \cite{Eliashberg,EliashbergInLangenbergLargkin1986}.

To accomplish this, we use the Keldysh formalism~\cite{LevchenkoPRB2007,Kamenev2011}.
In this case, besides the standard advanced and retarded Green's functions,
$G^{A}$ and $G^{R}$, respectively, one needs to include also the
Keldysh Green's function $G^{K}$. In situations close to equilibrium,
as it is our case, the latter is related to the former by $G^{K}=f(G^{R}-G^{A})$,
where $f$ is the symmetrized non-equilibrium distribution function,
$f(\xi)=1-2n_{F}(\xi)$. Within the Keldysh formalism, the quadratic
and quartic terms of the semi-classical energy functional Eq.~\eqref{EqGinzburgLandauForm}
must be rewritten in terms of classical and quantum source components
as $M^{cl}M^{q}$ ($\Delta^{cl}\Delta^{q}$) and $M^{cl}M^{cl}M^{cl}M^{q}$
($\Delta^{cl}\Delta^{cl}\Delta^{cl}\Delta^{q}$) in order for the
classical saddle point equations to be obtained via the constraints
on the action $\delta S/\delta M^{q}|_{M^{q}=0}=0$ and $\delta S/\delta\Delta^{q}|_{\Delta^{q}=0}=0$.
For the classical saddle point the RKA-rule applies, which implies
the following causal combination of the Green's functions in the quartic
coefficients: $RRRK+RRKA+RKAA+KAAA$.

Using $\tau$-matrices for the band space and $\sigma$ for the particle-hole
space, the relevant Green's functions are expressed as: 
\begin{equation}
\mathds{G}_{j,\alpha}^{R(K,A)}=\sum_{j,\alpha}\mathds{P}_{j,\alpha}G_{j,\alpha}^{R(K,A)}
\end{equation}
where $G_{j,\alpha}^{R(A)}=\operatorname{Lim}\limits _{\delta\rightarrow0^{+}}(\epsilon-\alpha\xi_{j}\pm i\delta)^{-1}$,
$\mathds{P}_{j,\alpha}=(1+j\tau_{z})(1+\alpha\sigma_{z})/4$ and $j,\alpha\in\{+,-\}$.
For convenience of notation, hereafter the hole-like band is associated
with $j=-1$ and the electron-like band is associated with $j=+1$.
The resulting Ginzburg-Landau coefficients are then given by: \begin{subequations}
\begin{align}
 & a_{m}=V_{\mathrm{SDW}}-2i\mathrm{Tr}\left[\mathds{G}_{-j,\alpha}^{\mu}\tau_{x}\mathds{G}_{j,\alpha}^{\nu}\tau_{x}\right]\label{EqCoefficientsDef}\\
 & a_{\Delta}=V_{\mathrm{SC}}-2i\mathrm{Tr}\left[\mathds{G}_{j,\alpha}^{\mu}\sigma_{x}\mathds{G}_{j,-\alpha}^{\nu}\sigma_{x}\right]\\
 & u_{m}=-2i\mathrm{Tr}\left[\mathds{G}_{-j,\alpha}^{\mu}\tau_{x}\mathds{G}_{j,\alpha}^{\nu}\tau_{x}\mathds{G}_{-j,\alpha}^{\beta}\tau_{x}\mathds{G}_{j,\alpha}^{\lambda}\tau_{x}\right]\\
 & u_{\Delta_{i}}=-2i\mathrm{Tr}\left[\mathds{G}_{j,\alpha}^{\mu}\sigma_{x}\mathds{G}_{j,-\alpha}^{\nu}\sigma_{x}\mathds{G}_{j,\alpha}^{\beta}\sigma_{x}\mathds{G}_{j,\alpha}^{\lambda}\sigma_{x}\right]\\
 & \gamma=2i\left\{ \mathrm{Tr}\left[\mathds{G}_{j,\alpha}^{\mu}\sigma_{x}\mathds{G}_{j,-\alpha}^{\nu}\sigma_{x}\mathds{G}_{j,\alpha}^{\beta}\tau_{x}\mathds{G}_{-j,\alpha}^{\lambda}\tau_{x}\right]\right.\nonumber \\
 & -\left.\mathrm{Tr}\left[\mathds{G}_{j,\alpha}^{\mu}\sigma_{x}\mathds{G}_{j,-\alpha}^{\nu}\tau_{x}\mathds{G}_{-j,-\alpha}^{\beta}\sigma_{x}\mathds{G}_{-j,\alpha}^{\lambda}\tau_{x}\right]\right\} 
\end{align}
\end{subequations} In all these expressions, the trace $\mathrm{Tr}[\ldots]$
implies summation over all indices, namely $\mathrm{Tr}[\ldots]=\sum_{j,\alpha}\int\frac{\mathrm{d}\omega}{2\pi}\int\left(\mathrm{d}p\right)\mathrm{tr}[\ldots]$,
as well as the Keldysh indices (Greek letters). The summation over
the latter satisfies the RKA rule, which means that for $(\mu,\nu)$,
only $(R,K)$ and $(K,A)$ are involved.

In order to evaluate the quartic coefficients, it is useful to consider
the generic combination of four Green's functions: 
\begin{equation}
\Gamma=2i\int\frac{\mathrm{d}\omega}{2\pi}\int\left(\mathrm{d}p\right)\mathrm{tr}\left[\mathds{P}_{a}\mathds{P}_{b}\mathds{P}_{c}\mathds{P}_{d}\right]\!\!\!\sum_{\substack{\mu\nu\beta\lambda\\
\mathrm{RKA-Rule}
}
}\!\!\!G_{a}^{\mu}G_{b}^{\nu}G_{c}^{\beta}G_{d}^{\lambda}
\end{equation}

\begin{figure}
\includegraphics[width=0.9\columnwidth]{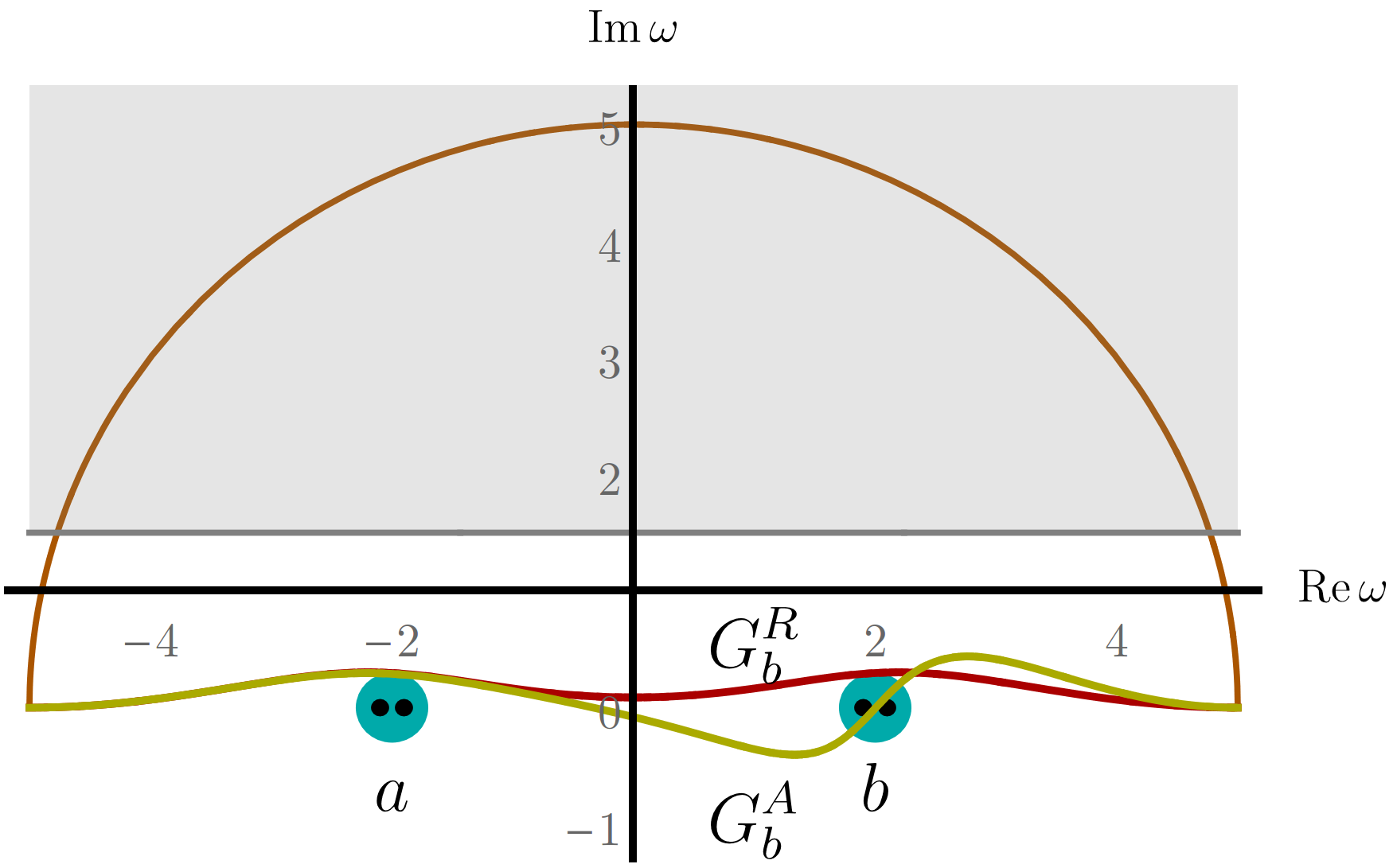}
\caption{Illustration of the contour integration that allows to extract the
relations used to determine the out of equilibrium coefficients in
Eq.~\eqref{EqIdentityExample}. The gray area contains the features
of the unknown function $f$ and the turquoise area corresponds to
the degenerate pole, which might only be resolved by causality. As
a consequence, the difference between retarded and advanced components
(as typically appearing from the Keldysh Green's function) is given
by the residue of single poles or multiple poles.}
\label{EqOutOfEqRelDerivation} 
\end{figure}

Here, the Latin letters correspond to the combined band and particle-hole
indices, whereas the Greek letters refer to Keldysh indices. Note
that due to the RKA-rule, only one of the Green's functions is a Keldysh
Green's function, as explained above. As a result, only the residue
of the Keldysh Green's function matters, since the Keldysh component
is strongly peaked at the Fermi level, $\operatorname{Lim}\limits _{\delta\rightarrow0}G_{a}^{R}-G_{a}^{A}=-2\pi i\delta(\epsilon-\xi_{a})$.
If, as in our case, two of the Greens functions share the same band/particle-hole
index $a$, a simple evaluation by means of a Dirac delta function
is no longer possible. Instead, one can use the identities: \begin{subequations}
\begin{multline}
\operatorname{Lim}\limits _{\delta\rightarrow0}2i\int\frac{\mathrm{d}\omega}{2\pi}G_{a}^{R}G_{b}^{R}G_{a}^{R}G_{b}^{K}=\\
2\Re\operatorname{Lim}\limits _{\delta\rightarrow0}\operatorname{Res}\limits _{G_{b}^{A},\mathrm{pole}}\left(G_{a}^{R}G_{b}^{R}G_{a}^{R}G_{b}^{A}f_{b}(\omega)\right),\label{EqIdentityExample}
\end{multline}
\begin{multline}
\operatorname{Lim}\limits _{\delta\rightarrow0}2i\int\frac{\mathrm{d}\omega}{2\pi}G_{a}^{K}G_{b}^{A}G_{a}^{A}G_{b}^{A}=\\
2\Re\operatorname{Lim}\limits _{\delta\rightarrow0}\operatorname{Res}\limits _{G_{a}^{R},\mathrm{pole}}\left(G_{a}^{R}G_{b}^{A}G_{a}^{A}G_{b}^{A}f_{a}(\omega)\right)
\end{multline}
\end{subequations} and for the causality-quenched configuration,
accordingly: \begin{subequations} 
\begin{align}
\operatorname{Lim}\limits _{\delta\rightarrow0} & 2i\int\frac{\mathrm{d}\omega}{2\pi}G_{a}^{R}G_{b}^{R}G_{a}^{K}G_{b}^{A}=\nonumber \\
2\Re & \operatorname{Lim}\limits _{\delta\rightarrow0}\left[\operatorname{Res}\limits _{G_{a}^{A},\mathrm{pole}}\left(G_{a}^{R}G_{b}^{R}G_{a}^{A}G_{b}^{A}f_{a}(\omega)\right)\right.\nonumber \\
 & +\left.\operatorname{Res}\limits _{G_{b}^{R},\mathrm{pole}}\left(G_{a}^{R}G_{b}^{R}G_{a}^{R}G_{b}^{A}f_{a}(\omega)\right)\right.\nonumber \\
 & -\left.\operatorname{Res}\limits _{G_{b}^{A},\mathrm{pole}}\left(G_{a}^{R}G_{b}^{R}G_{a}^{A}G_{b}^{A}f_{a}(\omega)\right)\right],
\end{align}
\begin{align}
\operatorname{Lim}\limits _{\delta\rightarrow0} & 2i\int\frac{\mathrm{d}\omega}{2\pi}G_{a}^{R}G_{b}^{K}G_{a}^{A}G_{b}^{A}=\nonumber \\
2\Re & \operatorname{Lim}\limits _{\delta\rightarrow0}\left[\operatorname{Res}\limits _{G_{b}^{R},\mathrm{pole}}\left(G_{a}^{R}G_{b}^{R}G_{a}^{A}G_{b}^{A}f_{b}(\omega)\right)\right.\nonumber \\
 & -\left.\operatorname{Res}\limits _{G_{a}^{R},\mathrm{pole}}\left(G_{a}^{R}G_{b}^{A}G_{a}^{A}G_{b}^{A}f_{b}(\omega)\right)\right.\nonumber \\
 & +\left.\operatorname{Res}\limits _{G_{a}^{A},\mathrm{pole}}\left(G_{a}^{R}G_{b}^{R}G_{a}^{A}G_{b}^{A}f_{b}(\omega)\right)\right],
\end{align}
\end{subequations} which are derived using methods of contour integration
as illustrated in Fig.~\ref{EqOutOfEqRelDerivation}.

As a consequence, we find the Ginzburg-Landau coefficients: \begin{subequations}
\label{EqCoefficients} 
\begin{align}
a_{m} & =V_{\mathrm{SDW}}^{-1}-2\int\left(\mathrm{d}p\right)\frac{f_{\bar{1}}-f_{1}}{\xi_{\bar{1}}-\xi_{1}}\\
a_{\Delta} & =V_{\mathrm{SC}}^{-1}-\int\left(\mathrm{d}p\right)\sum_{i}\frac{f_{i}}{\xi_{i}}\\
u_{m} & =-\int\left(\mathrm{d}p\right)\frac{2}{\left(\xi_{\bar{1}}-\xi_{1}\right)^{2}}\left[f_{\bar{1}}'+f_{1}'-2\,\frac{f_{\bar{1}}-f_{1}}{\xi_{\bar{1}}-\xi_{1}}\right]\\
u_{\Delta} & =-\frac{1}{2}\int\left(\mathrm{d}p\right)\sum_{i}\frac{1}{\xi_{i}^{2}}\left[f_{i}'-\frac{f_{i}}{\xi_{i}}\right]\\
\gamma & =\int\left(\mathrm{d}p\right)\!\frac{1}{\xi_{\bar{1}}-\xi_{1}}\left[\frac{f_{\bar{1}}}{\xi_{\bar{1}}^{2}}-\frac{f_{1}}{\xi_{1}^{2}}-\frac{f_{\bar{1}}'}{\xi_{\bar{1}}}+\frac{f_{1}'}{\xi_{1}}\right]
\end{align}
\end{subequations} where we used $\bar{1}=-1$, $f_{i}=f(\xi_{i})$,
$f'=\partial_{\xi}f(\xi)$ and $\int\left(\mathrm{d}p\right)=\int\frac{d^{2}p}{4\pi^{2}}$.
In equilibrium, where $f\left(\xi\right)=\tanh\left(\frac{\beta\xi}{2}\right)$,
the expressions reduce to those derived in Ref. \cite{FernandesSchmalianPRB2010}.

Because the two band dispersions are parametrized by $\xi_{1}\equiv-\xi-b\left(\theta\right)$
and $\xi_{\bar{1}}=\xi$, with $b\left(\theta\right)=\delta_{\mu}+\delta_{m}\cos2\theta$,
the momentum integration can be split into an integration over momentum
perpendicular to the FS ($\xi$) and momentum parallel to the FS ($\theta$),
$\int\left(\mathrm{d}p\right)=N_{F}\int\frac{\mathrm{d}\theta}{2\pi}\int\mathrm{d}\xi$,
where $N_{F}$ is the density of states. As a result, all the integrals
over $\xi$ can be performed analytically, leaving only the angular
integrals to be evaluated numerically. In terms of the Fourier components
of the anisotropic temperature $\delta T_{n}$, we find: \begin{subequations}
\label{EqLGCoeffNotion} 
\begin{align}
a_{m} & =4N_{F}\left(\log\frac{T}{T_{M}}-\hat{a}_{m,\mathrm{eq}}+\sum_{n}\delta\hat{a}_{m,n}\frac{\delta T_{n}}{T}\right)\\
a_{\Delta} & =4N_{F}\left(\log\frac{T}{T_{c}}+\sum_{n}\delta\hat{a}_{\Delta,n}\frac{\delta T_{n}}{T}\right)\\
u_{m} & =\frac{N_{F}}{T^{2}}\left(\hat{u}_{m,\mathrm{eq}}+\sum_{n}\delta\hat{u}_{m,n}\frac{\delta T_{n}}{T}\right)\\
u_{\Delta} & =\frac{N_{F}}{T^{2}}\left(\hat{u}_{\Delta,\mathrm{eq}}+\sum_{n}\delta\hat{u}_{\Delta,n}\frac{\delta T_{n}}{T}\right)\\
\gamma & =\frac{N_{F}}{T^{2}}\left(\hat{\gamma}_{\mathrm{eq}}+\sum_{n}\delta\hat{\gamma}_{n}\frac{\delta T_{n}}{T}\right)
\end{align}
\end{subequations} The equilibrium coefficients are given by: \begin{subequations}
\begin{align}
\hat{a}_{m,\mathrm{eq}} & =2(\gamma_{E}+\log4)+\sum_{\alpha=\pm1}\left\langle \psi^{(0)}(X_{\alpha})\right\rangle \label{EqEqFunctions}\\
\hat{u}_{m,\mathrm{eq}} & =-\frac{1}{8\pi^{2}}\sum_{\alpha=\pm1}\left\langle \psi^{(2)}(X_{\alpha})\right\rangle \\
\hat{u}_{\Delta,\mathrm{eq}} & =\frac{7}{2\pi^{2}}\zeta(3)\\
\hat{\gamma}_{\mathrm{eq}} & =\left\langle \!\frac{\psi^{(0)}\left(\frac{1}{2}\right)}{2\pi^{2}x^{2}}\!\right\rangle -\sum_{\alpha=\pm1}\left\langle \!\frac{\psi^{(0)}(X_{\alpha})-i\alpha\frac{x}{\pi}\psi^{(1)}(X_{\alpha})}{(2\pi x)^{2}}\!\right\rangle 
\end{align}
\end{subequations} where we introduced the notation $x=-\frac{\delta_{\mu}}{\pi T}-\frac{\delta_{m}}{\pi T}\cos2\theta$
and $X_{\alpha}=\frac{1}{2}+\alpha ix$; the brackets denote integration
over the angular variable and $\psi^{(n)}$ is the polygamma function
of order $n$. The non-equilibrium coefficients are given by: \begin{subequations}
\begin{align}
\delta\hat{a}_{m,n}(x) & =1-\left\langle \frac{ix}{2}\sum_{\alpha=\pm1}\alpha\psi^{(1)}(X_{\alpha})\,\cos n\theta\right\rangle \label{EqNonEqFunctions}\\
\delta\hat{a}_{\Delta,n} & =\left\langle \cos n\theta\right\rangle =\delta_{n,0}\\
\delta\hat{u}_{m,n} & =-\frac{1}{32\pi^{2}}\times\\
 & \times\sum_{\alpha=\pm1}\left\langle \left[\psi^{(2)}(X_{\alpha})+\frac{\alpha ix}{2}\psi^{(3)}(X_{\alpha})\right]\cos n\theta\right\rangle \\
\delta\hat{u}_{\Delta,n} & =\frac{7}{2\pi^{2}}\zeta(3)\left\langle \cos n\theta\right\rangle =\frac{7}{2\pi^{2}}\zeta(3)\delta_{n,0}\\
\delta\hat{\gamma}_{n} & =-\frac{1}{8\pi^{2}}\sum_{\alpha=\pm1}\left\langle \psi^{(2)}(X_{\alpha})\,\cos n\theta\right\rangle 
\end{align}
\end{subequations} Note that, in the spirit of the Ginzburg-Landau
approach, the temperature in the pre-factors of the quartic coefficients
must be replaced by the temperature at which both transition lines
meet, $T_{M}=T_{c}$. We verified that the equilibrium phase diagram
resulting from these equations reproduces very well the transition
lines of the phase diagram of Ref. \cite{VorontsovVavilovPRB2010}
(including the $T_{M}$ and $T_{c}$ lines below, but in the vicinity
of, the multicritical point), which used the full non-linear gap equations.

\section{Impact of other scattering sources \label{app_impurity}}

As discussed in the main text, there are two additional sources of
scattering that may impact our results: impurity scattering and electron-electron
scattering. These contributions can be included in the Bolztmann equation
via their collision integrals:
\begin{widetext}
\begin{align}
I_{\mathrm{coll}}^{\mathrm{imp}}[n^{F}]=-\int\frac{\mathrm{d}^{2}p'}{(2\pi)^{2}}\delta(\xi_{\vec{p}}-\xi_{\vec{p}'})|M_{\vec{p}-\vec{p}'}^{\mathrm{imp}}|^{2}\left[n_{\xi_{\vec{p}}}^{F}-n_{\xi_{\vec{p}'}}^{F}\right] &  & \hspace*{7.3cm}\label{EqImpScatteringKernel}
\end{align}
\begin{multline}
I_{\mathrm{coll}}^{\mathrm{el-el}}[n^{F}]=-\int\frac{\mathrm{d}^{2}p'}{(2\pi)^{2}}\int\frac{\mathrm{d}^{2}\tilde{p}}{(2\pi)^{2}}\int\frac{\mathrm{d}^{2}\tilde{p}'}{(2\pi)^{2}}\delta(\xi_{\vec{p}}-\xi_{\vec{p}'}+\xi_{\tilde{\vec{p}}}-\xi_{\tilde{\vec{p}}'})\delta(\vec{p}-\vec{p}'+\tilde{\vec{p}}-\tilde{\vec{p}}')|M_{\vec{q}}^{\mathrm{el-el}}|^{2}\times\\
\times\left[n_{\xi_{\vec{p}}}^{F}n_{-\xi_{\vec{p}'}}^{F}n_{\xi_{\tilde{\vec{p}}}}^{F}n_{-\xi_{\tilde{\vec{p}}'}}^{F}-n_{-\xi_{\vec{p}}}^{F}n_{\xi_{\vec{p}'}}^{F}n_{-\xi_{\tilde{\vec{p}}}}^{F}n_{\xi_{\tilde{\vec{p}}'}}^{F}\right]\label{EqElElScatteringKernel}
\end{multline}
where $|M_{\vec{p}-\vec{p}'}^{\mathrm{imp}}|^{2}\nu_{F}=\tau_{\mathrm{imp}}^{-1}$.
To understand how these additional contributions affect the solution
of the Boltzmann equation that we found, it is instructive to consider
the functional form of the latter, as we did in Eq. \eqref{EqFunctionalAbsorbtionEmmisionSummationn}.
For convenience, we rewrite the functional in terms of a kernel function
$K\left(h^{F}\right)$
\end{widetext}

\begin{equation}
\mathcal{F}\left[h^{F}\right]=\iint\frac{\mathrm{d}\theta\mathrm{d}\theta'}{(2\pi)^{2}}N_{F}(\theta)N_{F}(\theta')\frac{\omega_{q}|M_{\vec{q}}|^{2}}{4\sinh^{2}(\frac{\beta\omega_{q}}{2})}\,K\left(h^{F}\right)\label{general_F_K}
\end{equation}

When only non-equilibrium acoustic phonons are present, the kernel
is given by:

\begin{align}
K_{\mathrm{phonon}}\left(h^{F}\right) & =\left(\frac{\beta\omega_{q}}{2}\right)^{2}\left[\left(\delta T_{\hat{\vec{p}}}+\delta T_{\hat{\vec{p}}'}\right)-\tilde{h}_{q}^{B}\right]^{2}\nonumber \\
 & +\frac{1}{3}\left[\pi^{2}+\left(\frac{\beta\omega_{q}}{2}\right)^{2}\right]\left(\delta T_{\hat{\vec{p}}}-\delta T_{\hat{\vec{p}}'}\right)^{2}\label{Phonon_Kernel}
\end{align}
where $\tilde{h}_{q}^{B}=2h_{q}^{B}/\beta\omega_{q}$ is proportional
to the driving term. This expression is equivalent to Eq. \eqref{EqFunctionalAbsorbtionEmmisionSummationn},
since the quadratic term in $\tilde{h}^{B}$ just provides a trivial
shift of the functional. 

When expressed in this form, the functional reveals in a very transparent
way the competition between two opposing effects, represented by the
two positive-defined terms of the kernel. The first term, arising
from the driving of the acoustic phonons, is minimized by setting
an anisotropic effective temperature profile $\left(\delta T_{\hat{\vec{p}}}+\delta T_{\hat{\vec{p}}'}\right)=\tilde{h}_{q}^{B}$.
The second term, on the other hand, corresponds to the relaxation
of the electrons back to a uniform temperature profile, since this
term is minimized by $\delta T_{\hat{\vec{p}}}-\delta T_{\hat{\vec{p}}'}=0$.
By comparing the coefficients of this term, it is clear that as long
as $\frac{\beta\omega_{q}}{2}\gtrsim\pi$, both terms are comparable.
In the case where $\frac{\beta\omega_{q}}{2}\lesssim\pi$, the second
term dominates, and the effective temperature profile is expected
to become less anisotropic. This is precisely what we note from our
numerical results shown in Fig.~\ref{FigNumericallyDeterminedTemperatureProfile}c
of the main text: when $\Omega_{0}\equiv\frac{\beta\omega_{0}}{2}$
decreases towards $1$, the ratio $\delta T(\frac{\pi}{4})/\delta T(0)$
also decreases.

This qualitative analysis can be extended to include the effects of
impurity scattering and electron-electron scattering. The impurity
collision integral shown in Eq. \eqref{EqImpScatteringKernel} gives
rise to the following functional kernel:

\begin{equation}
K_{\mathrm{imp}}\left(h^{F}\right)=\frac{|M^{\mathrm{imp}}|^{2}}{|M(\hat{\vec{q}})|^{2}}\frac{4\sinh^{2}(\beta\omega_{q}/2)}{\beta\omega_{q}}\frac{\pi^{2}}{3}\left(\delta T_{\hat{\vec{p}}}-\delta T_{\hat{\vec{p}}'}\right)^{2}
\end{equation}

It is clear that impurity scattering favors a uniform effective temperature
profile, adding up to the second term of the phonon kernel in Eq.
\eqref{Phonon_Kernel}. Thus, for the first term of $K_{\mathrm{phonon}}$
to be comparable to these terms, the following condition has to be
met:

\begin{equation}
\left(\frac{\beta\omega_{q}}{2}\right)^{2}\gtrsim\pi^{2}+4\pi^{2}\frac{|M^{\mathrm{imp}}|^{2}}{|M(\hat{\vec{q}})|^{2}}\frac{\sinh^{2}(\beta\omega_{q}/2)}{\beta\omega_{q}}\label{EqImpScatt}
\end{equation}

Thus, for temperatures comparable to the energy of the excited phonon
mode, $T\sim\omega_{0}$, the anisotropic temperature profile resulting
from the minimization of the functional would persist even if the
impurity and phonon matrix elements are of the same order. Of course,
as temperature becomes much smaller than $\omega_{0}$, the effect
will only persist if $\left|M^{\mathrm{imp}}\right|^{2}\ll\left|M(\hat{\vec{q}})\right|^{2}$.
There are basically two different scenarios in which this condition
is satisfied. The first is when the system is clean, since $|M^{\mathrm{imp}}|^{2}$
scales with the impurity concentration. The second is when the sound
velocity $v_{s}$ of the system is small, which makes $\left|M(\hat{\vec{q}})\right|^{2}$
large.

The impact of the electron-electron scattering, described by the collision
integral \eqref{EqElElScatteringKernel}, is more subtle. First, one
has to distinguish energy relaxation and momentum relaxation processes.
As discussed in the main text, the former is expected to be faster
than the latter, particularly in two-dimensional systems. It is the
energy relaxation that leads to a rather quick thermalization of the
electronic degrees of freedom; however, this process does not smear
out the anisotropic effective temperature caused by the non-equilibrium
phonons, as it is not capable of relaxing momentum. On the contrary,
this process actually helps to establish the anisotropic effective
temperature, as it enforces the electronic distribution to be nearly
thermal, as we tacitly assumed when we linearized the Boltzmann equation. 

Momentum relaxation due to electron-electron scattering, on the other
hand, has the potential to wash away the anisotropy in the temperature.
To estimate this effect, we recast the collision integral Eq.~\eqref{EqElElScatteringKernel}
in a functional form. The form of the functional, however, is different
than that of Eq.~\eqref{general_F_K}, because it involves four electronic
states. Generally, the functional is given by
\begin{widetext}
\begin{multline}
\iint\frac{\mathrm{d}\theta\mathrm{d}\theta'}{(2\pi)^{2}}N(\theta)N(\theta')\iint\frac{\mathrm{d}\tilde{\theta}\mathrm{d}\tilde{\theta}'}{(2\pi)^{2}}N(\tilde{\theta})N(\tilde{\theta}')\frac{(\beta\omega_{q})^{2}|M_{\vec{q}}^{\mathrm{el-el}}|^{2}}{4\sinh^{2}(\beta\omega_{q}/2)}\left[\frac{1}{3}\left(\pi^{2}+\frac{1}{2}\beta^{2}\omega_{q}^{2}\right)\left(\left(\delta T_{\hat{\vec{p}}}-\delta T_{\hat{\vec{p}}'}\right)^{2}+\left(\delta T_{\hat{\tilde{\vec{p}}}}-\delta T_{\hat{\tilde{\vec{p}}}'}\right)^{2}\right)\right.\\
\left.+\left(\frac{\beta\omega_{q}}{2}\right)^{2}\left(\left(\delta T_{\hat{\vec{p}}}-\delta T_{\hat{\tilde{\vec{p}}}}\right)^{2}+\left(\delta T_{\hat{\vec{p}}}-\delta T_{\hat{\tilde{\vec{p}}}'}\right)^{2}+\left(\delta T_{\hat{\vec{p}}'}-\delta T_{\hat{\tilde{\vec{p}}}}\right)^{2}+\left(\delta T_{\hat{\vec{p}}'}-\delta T_{\hat{\tilde{\vec{p}}}'}\right)^{2}\right)\right]\label{EqCoulombScat}
\end{multline}
\end{widetext}

If we attempt to recast this functional in the form given by Eq.~\eqref{general_F_K},
we have to integrate out the momenta $\tilde{\theta}$ and $\tilde{\theta}'$,
which will generate non-linear effects. For simplicity, if we focus
only on the first term, and cast it in the form of Eq. \ref{general_F_K},
we find the following kernel: 

\begin{align}
K^{\mathrm{el-el}}\left(h^{F}\right) & =\frac{|M_{\vec{q}}^{\mathrm{el-el}}|^{2}}{|M(\hat{\vec{q}})|^{2}}\nu_{F}^{2}\beta\omega_{q}\nonumber \\
 & \times\frac{1}{3}\left(\pi^{2}+\frac{1}{2}\beta^{2}\omega_{q}^{2}\right)\left(\delta T_{\hat{\vec{p}}}-\delta T_{\hat{\vec{p}}'}\right)^{2}
\end{align}

Thus, for $T\sim\omega_{0}$, this term does not significantly enhance
the corresponding term of the phononic kernel, Eq. \eqref{Phonon_Kernel},
provided that $|M_{\vec{q}}^{\mathrm{el-el}}|^{2}\nu_{F}^{2}$ is
of the same order as $|M(\hat{\vec{q}})|^{2}$. 
As in the case of impurities, the phononic contribution $|M(\hat{\vec{q}})|^{2}$
will be enhanced in systems near a structural transition.
\end{document}